\documentclass{elsarticle}

\usepackage{adjustbox}
\usepackage{array}
\usepackage{booktabs}
\usepackage{multirow}
\usepackage{xspace}
\usepackage{amsfonts} 
\usepackage{amsmath}
\usepackage{amssymb}
\usepackage{forest}
\usepackage{pifont}
\usepackage{diagbox}
\usepackage{subfigure} 

\usepackage{mathtools}
\usepackage{stmaryrd}
 
\usepackage[utf8]{inputenc} 
\usepackage{natbib} 
\usepackage{forest}
\usepackage{amsmath}
\usepackage{tabularx}
\usepackage{lscape}

\usepackage{ntheorem}

\makeatletter
\renewtheoremstyle{plain}
  {\item{\theorem@headerfont ##1\ ##2\theorem@separator}~}
  {\item{\theorem@headerfont ##1\ ##2\ (##3)\theorem@separator}~}
\makeatother

{\theoremheaderfont{\upshape\bfseries}
 \theorembodyfont{\normalfont\slshape}
 \newtheorem{definition}{Definition}
 \newtheorem{metadefinition}{Meta-Definition}}

\def\Y{\textbf{Yes}\xspace}
\def\N{\textbf{No}\xspace}

\newcommand{\powerset}{\raisebox{.15\baselineskip}{\Large\ensuremath{\wp}}}

\begin{document}

 \title{Multilayer network simplification:\\ approaches, models and methods}

\author[1]{Roberto Interdonato}
\ead{roberto.interdonato@cirad.fr}

\author[2]{Matteo Magnani}
\ead{matteo.magnani@it.uu.se}

\author[3]{Diego Perna}
\ead{d.perna@dimes.unical.it}

\author[3]{Andrea Tagarelli\corref{cor1}}
\ead{andrea.tagarelli@unical.it}

\author[2]{Davide Vega}
\ead{davide.vega@it.uu.se}

\cortext[cor1]{Corresponding author}

\address[1]{CIRAD, UMR TETIS, F-34398 Montpellier, France. \\TETIS, Univ Montpellier, AgroParisTech, CIRAD, CNRS, INRAE, Montpellier, France}
\address[2]{InfoLab, Dept. of Information Technology, Uppsala University, Sweden}
\address[3]{Dept. of Computer Engineering, Modeling, Electronics, and Systems Engineering (DIMES), University of Calabria, Rende (CS), Italy}

\begin{abstract}
Multilayer networks have been widely used to represent and analyze systems of interconnected entities where both the entities and their connections can be of different types. However, real multilayer networks can be difficult to analyze because of irrelevant information, such as layers not related to the objective of the analysis, because of their size, or because traditional methods defined to analyze simple networks do not have a straightforward extension able to handle multiple layers. Therefore, a number of methods have been devised in the literature to \emph{simplify} multilayer networks with the objective of improving our ability to analyze them. In this article we provide a unified and practical taxonomy of existing simplification approaches, and we identify categories of multilayer network simplification methods that are still underdeveloped, as well as emerging trends.
\end{abstract}

\maketitle

\section{Introduction}
\label{sec:intro}

The network analysis and mining research field has raised in popularity in the last two decades, thanks to the ability of networks of representing a wide range of real-life phenomena from physical to biological and social systems, from scientific to financial data, transportation routes, and many more.  
In this regard, the \textit{multilayer network model} is widely used as a powerful tool to represent the organization and relationships of complex data in many domains. Multilayer networks, which initially gained momentum in social computing~\cite{Dickison2016}, are designed to provide a more realistic representation of the different and heterogeneous relations that may characterize   an entity in the network system. For instance, a multilayer network enables an expressive way to   model different types of social relations among the same set of individuals, where layers correspond to  different  on-line as well as off-line relations (e.g., following, co-authorship, co-working relations,  and so on).

However, as we already witnessed at the beginning of the data mining era,  the availability of huge amounts of complex  network data represents an invaluable potential but also inevitably leads to processing issues. Just think of the number of  monthly active users for the main online social networks, which is, at the time of writing, 
around 335 millions for Twitter\footnote{https://www.statista.com/statistics/282087/number-of-monthly-active-twitter-users/} and more than 2 billions for Facebook\footnote{https://www.statista.com/statistics/264810/number-of-monthly-active-facebook-users-worldwide/}. Modeling these networks in their entirety for analysis purposes becomes  unfeasible in most cases, and focusing on limited portions of the network (e.g., related to  specific phenomena or geographical areas) is likely to   cause problems in the \textit{boundary specification}~\cite{lauman,Kossinets06}, i.e., the choice of which entities and relations should be included in the data.
Moreover, when dealing with  multilayer networks, the boundary specification problem is even amplified: in fact, we can recognize  a  \textit{horizontal} boundary specification problem for each layer similar to the one observed for single-layer networks, that is, the choice of which actors to include in the network, and a \textit{vertical} boundary specification problem~\cite{Dickison2016}, i.e., the problem of choosing which types of relations should be represented in the network (i.e., how many layers and with which semantics).  
Given these premises, it is easy to understand how most network data modeled upon real-world phenomena may be \textit{incomplete} and/or \textit{noisy}: in fact,  relations that are supposed to be  central for a specific analysis task may be missing, or hidden under a considerable amount  of irrelevant information. In certain cases, the existence of the relations and their strength may not even be possible to determine with certainty, leading to probabilistic representations~\cite{ParchasGPB15}.

Several   network  processing techniques have been proposed   to partially overcome the above problems in order to enable complex analysis tasks on very large networks.  
Our goal in this work is to bring order to the existing literature on approaches, models and methods for \textit{simplification tasks in  multilayer networks}. 
With the term ``simplification'' here we refer to a specific type of network manipulation  that aims at \textit{simplifying} the     structure of a network. 
 We deliberately utilize the term with quite a broad meaning, which anyway does not coincide, hence should not be confused, with the mechanism of mapping multiple edges to single edges and removing self-loops.  Rather, the choice of such a broad term derives from the observation that although a significantly large amount of techniques that may be described as  simplification ones have been proposed in the literature, most of them were designed to solve specific problems in different domains; by contrast,   nowadays we recognize a clear need to systematize these techniques in the context of complex network data, with emphasis on the multilayer network model, yet   regardless of the peculiarities of a particular application domain.  
Network simplification can be seen as a special case of \textit{manipulation}, which also includes other tasks such as \textit{perturbation} and \textit{refinement}. The former includes techniques designed for altering information encoded in a network, generally for privacy reasons (such as obfuscation and encryption techniques), whereas the latter refers to  methods that are conceived to infer missing relations or attributes, or to correct the information encoded in a network (e.g., based on ontological facts).  

We identify three broad categories of network simplification: \textit{selection}, \textit{aggregation}, and \textit{transformation}. 
  Selection   methods   operate on a multilayer network to reduce its size by filtering or sampling subsets of nodes, edges and/or layers, according to specific features of the entities involved or predefined model characteristics to preserve.  
 Aggregation   refers to various approaches to define partitional or hierarchical grouping mechanisms that involve nodes, edges or layers such as layer-based flattening, coarsening, summarization, community detection, and positional equivalence.   Transformation approaches are divided into \textit{projection} and \textit{graph embedding} methods. 
 Projection  methods are designed to   deal with   different node (entity) types in a network, and  aim to replace  nodes of selected types with relations.
 Finally, graph embedding  techniques aim to transform a graph into a  low-dimensional, vectorial representation, which is also key enabling for machine and deep learning tasks.

Motivations  for performing a simplification task on a multilayer network are   manifold and often they are raised from different requirements in the target application domain. In this regard, we can recognize  the following computational aspects for which a  network simplification task can be beneficial:   
\begin{itemize}
\item By solving  noise or incompleteness issues in a complex network, the relevant information contained in the network will more easily be unveiled, leading to   \textit{improved data quality}. This   is expected to  have a beneficial impact on the effectiveness of methods to be applied for further analysis tasks.
\item Simplifying a complex network can lead to \textit{improved  performance} of further  computational analysis methods which may struggle with  efficiently handling  very large networks.  
\item Simplifying a complex network can also \textit{enable application}  of an existing  method originally conceived for simple (i.e., monoplex) networks, or can aid to cope with  model compatibility issues when it is not possible to apply a selected method on a given network model. 
\end{itemize}

\textbf{Contributions.\ } 
In this work, we  provide the first conceptualization of  the network simplification problem for multilayer networks, for which we recognize and formally define three main categories. According to this classification, we propose  a  formal systematization of approaches, models and methods related to network simplification tasks. 
 
One major goal of our work is to discuss how simplification approaches  that were  conceived for simple networks could only be extended, adapted, or redefined to deal with multilayer networks. In this regard, whenever there is a lack in recent literature to support pursuing the above goal, we eventually try to hint at methodological solutions for specific classes of simplification techniques for multilayer networks.

\textbf{Limitations and scope.\ }  
In this work, we will focus on a topology-driven multilayer network model, therefore we will leave out of consideration techniques that are designed to deal with  node and/or edge attributes, such as  reduction of the number of attributes associated to nodes/edges in a network (e.g., feature selection methods), or  reduction of the cardinality of the value set for a certain attribute (e.g., discretization, binning).
 We consider the above techniques closer to a traditional data mining scenario than to a network mining one, and they are often domain-specific.    
More details on feature selection, discretization and other methods focusing on attribute values rather than network structure can be found in most data mining and machine learning textbooks.  

It should be noted that, although the focus of this work is on multilayer networks,   we will also discuss how simplification techniques that are not originally conceived for  multilayer networks can be applied to such networks. In this respect, we   refer the reader to more focused surveys that cover one or more  topics related to the ones discussed in this work but referring to   single-layer networks only. For instance, Liu et al.~\cite{LiuSDK18} overview  methodologies for   static and dynamic graph summarization, which can also support related tasks,  such as compression and clustering;  
 Beck et al.~\cite{BeckBDW17} provides a comprehensive survey on  visualization of dynamic graphs, which has also attracted increasing interest from different research communities.

\textbf{Plan of this paper.\ }
The rest of the paper is organized as follows. 
 We provide formal definitions for each of the three network simplification categories in Section~\ref{sec:problemstatement}. Accordingly, in Section~\ref{sec:literature} we classify existing methods in the literature in the context of our taxonomy and provide an overview of the main methods for each category, so that the readers can use this article to identify potentially useful approaches for their simplification problems. This overview of the literature allows us to identify categories of multilayer network simplification methods that are still underdeveloped, as well as emerging trends, as a starting point for future research. A forward-looking discussion of these and other general aspects emerging from our classification and literature review is presented in Section~\ref{sec:discussion}. 
  Moreover, we review the available software implementations of methods for multilayer network analysis with emphasis on network simplification. One of the objectives of this article is indeed to boost the integration of individual methods into more general libraries and frameworks, to make them more easily usable and extensible. 
  Finally, in Section~\ref{sec:future}, we sum up the limitations of existing methods for multilayer network simplification, and draw several  pointers for future research.

\section{Definitions of Network Simplification}
\label{sec:problemstatement}

Given a set of actors $\mathcal{A}$ and a set of
layers $\mathcal{L}$, a \textit{multilayer network} is defined as a quadruple $G = (\mathcal{A}, \mathcal{L}, V, E)$ where $(V, E)$ is a graph, $V \subseteq \mathcal{A} \times \mathcal{L}$ and $E \subseteq V \times V$.
 %
 Each actor must be present in at least one layer,  but each  layer is not required to contain all  actors. 
 Each node  in one layer could be  linked to nodes corresponding to the same actor in a few or all other layers;  in the \textit{multiplex} setting,    the inter-layer links only connect the same actor in different layers.     

In the following, we provide a meta-definition of the network simplification problem, 
which is meant to establish a formal backbone for all simplification tasks under consideration in this work, i.e., selection, aggregation, and transformation. 
 Table~\ref{tab:notation} shows the main notation introduced in this section.

\begin{table}[t!]
\centering
\caption{Main notation used in this paper.} 
\label{tab:notation}
\begin{tabular}{|cl|} 
\hline
\textit{notation} & \textit{description} \\ \hline \hline 
$G$ & multilayer network graph\\
$\mathcal{A}$ & set of actors \\
$L$; $\mathcal{L}$; $l$ & Layer; set of layers; number of layers \\
$V$ & set of nodes \\
$E$ & set of edges \\
$\powerset$; $p$ & power set  function; set function \\
$\theta$; $\Theta$ & expression; set of expressions \\
$\mathcal{T}$ & set of string constants\\
$f_\theta$ & simplification function\\
$V_f$ & flattened set of nodes\\
$E_f$ & flattened set of edges\\
 $\omega, w$ &   actor, resp. layer, weighting functions\\
$S$, $C$ & coarsened graph, set of corrections \\
$\mathbf{k}$ & coreness vector\\
$d$ & size of the embedding \\
$\mathbf{v}_n^i$ & embedding vector for node $n$ in layer $L_i$\\ 
\hline
\end{tabular} 
\end{table}

 A key constituent of our meta-definition corresponds to a function, denoted as  $f_\theta$, that describes the modification to a multilayer network carried out by a simplification method. Depending on the type of \textit{expressions} $\Theta$, the simplification function is further  designed to support a specific simplification task.  
 The outcome of a network simplification process is a multilayer network whereby one or more of the structural elements of the original network   (i.e., $\mathcal{A}$, $\mathcal{L}$, $V$, $E$) are being affected by the process.  
 Formally, this is denoted at the end of the meta-definition, as well as at the end of each of the subsequent specialized definitions,   by a disjunction of conditions that declare which  network elements are affected by a selection or modification of their respective sets,  
  for each simplification category; this information is also summarized and concisely reported in the column \textit{affected elements} of Table~\ref{tab:cat}, which will be discussed later in   Section~\ref{sec:literature}.

\begin{metadefinition}[Network Simplification]\label{def:ns}
Let  $\mathcal{G}(\mathcal{A}, \mathcal{L})$ denote the set of all possible multilayer networks that can be defined over $\mathcal{A}$ and $\mathcal{L}$. Moreover, let $\Theta(\mathcal{A}, \mathcal{L}, V, E)$ denote a set of expressions on alphabets  $\mathcal{A}, \mathcal{L}, V, E$.
 Given a multilayer network $G = (\mathcal{A}, \mathcal{L}, V, E) \in \mathcal{G}(\mathcal{A}, \mathcal{L})$, expressions $\theta \subseteq \Theta(\mathcal{A}, \mathcal{L}, V, E)$, let  $f_{\theta}:  \mathcal{G}(\mathcal{A}, \mathcal{L}),\theta  \mapsto  
  \mathcal{G}(p(\mathcal{A}), p(\mathcal{L}))$ be a \textit{network simplification} function,  such that the set of actors $p(\mathcal{A})$ and the set of layers $p(\mathcal{L})$ in the simplified network are defined as follows:   either $p(\mathcal{A})  \subseteq \mathcal{A}$  
  or $p(\mathcal{A}) \subset \powerset(\mathcal{A})$, 
  and  either $p(\mathcal{L}) \subseteq  \mathcal{L}$ or  $p(\mathcal{L}) \subset  \powerset(\mathcal{L})$, where $\powerset(\mathcal{A})$ and $\powerset(\mathcal{L})$ are the power sets of actors and layers, respectively.   
  The   network simplification problem is to obtain a   network $f_{\theta}(G) = G'=(\mathcal{A}', \mathcal{L}', V',E')$ such that the following disjunction of conditions holds: 
$|\mathcal{A}'| < |\mathcal{A}| \vee  |\mathcal{L}'| < |\mathcal{L}| \lor
|V'| < |V| \lor |E'| < |E|$.
\end{metadefinition}

\begin{figure}[t!]
\centering
\includegraphics[width=1\textwidth]{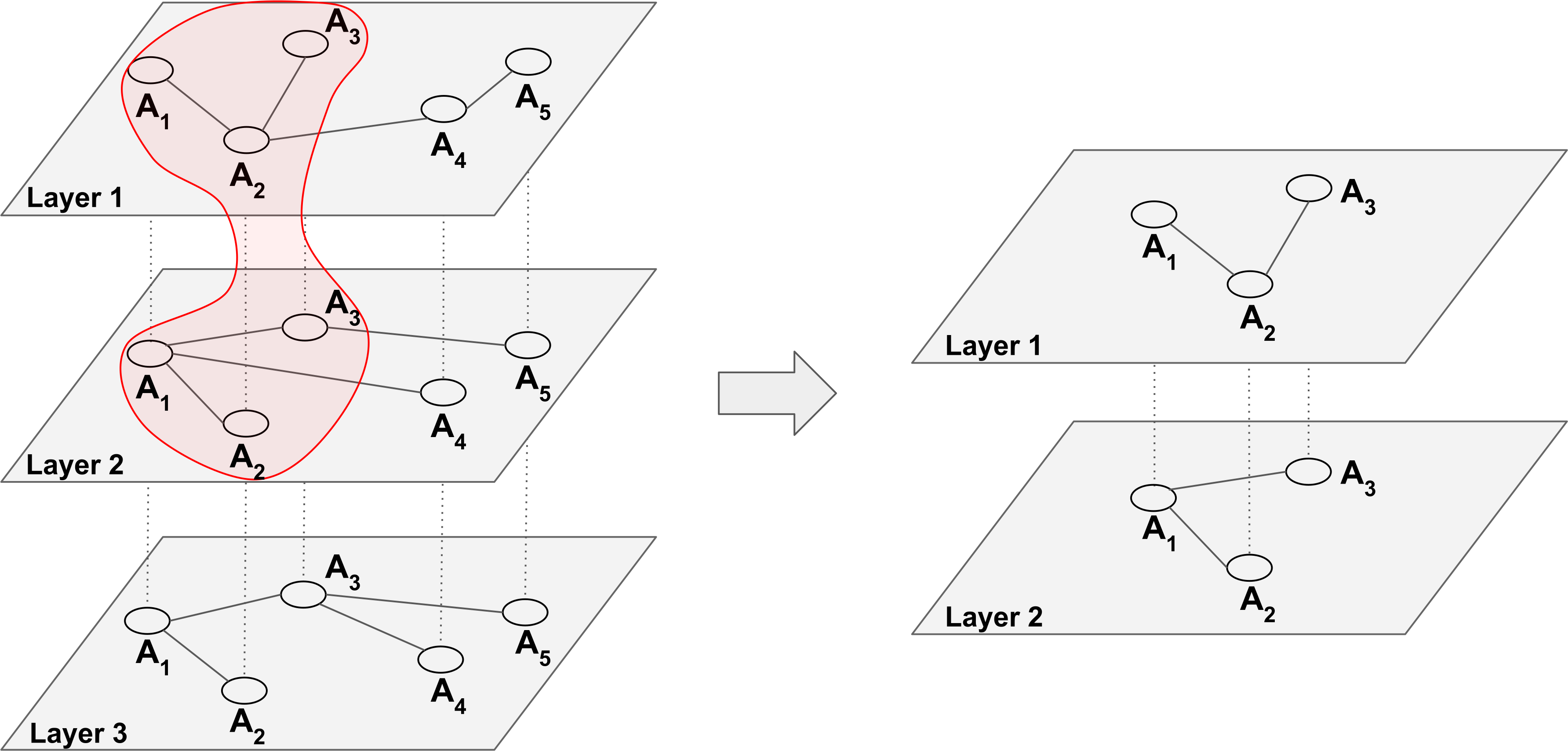}
\caption{Illustrative example of selection-oriented simplification: the subset of actors $\{A_1, A_2, A_3\}$ and the subset of layers $\{L_1, L_2\}$ are selected from  an input multilayer network (on the left).}
\label{fig:selection} 
\end{figure}

Definition~\ref{def:ns} is supposed to be specialized for each  of the three simplification categories. In the \textit{selection} case, 
the simplification function depends on \textsf{select} expressions, i.e., it is defined to identify a subset of   $\mathcal{A}$ and/or  $\mathcal{L}$,  and consequently $V$ and/or $E$, as provided in the following definition.  
 Figure~\ref{fig:selection} shows an example of selection-oriented simplification.

\begin{definition}[Selection-oriented Network Simplification]\label{def:selection_ns}
Given a multilayer network graph  $G = (\mathcal{A}, \mathcal{L}, V, E)$, let  $f_{\theta}$ be a simplification function with  $\theta \subseteq \Theta(\mathcal{A}, \mathcal{L}, V, E)$ such that $\Theta$ is a set of expressions of the form $\textsf{select}(X)$, with variables $X \in \{\mathcal{A}, \mathcal{L}, V, E\}$.   
The  \textit{Selection-oriented} network simplification problem  is to obtain a   network $G'=(\mathcal{A}', \mathcal{L}', V',E')$, where $G \xmapsto{f_{\theta}} G'$,     s.t. $
\mathcal{A}' \subset \mathcal{A} \lor \mathcal{L}' \subset \mathcal{L} \lor V' \subset V \lor E' \subset E $.
\end{definition}

\begin{figure}[t!]
\centering
\includegraphics[width=1\textwidth]{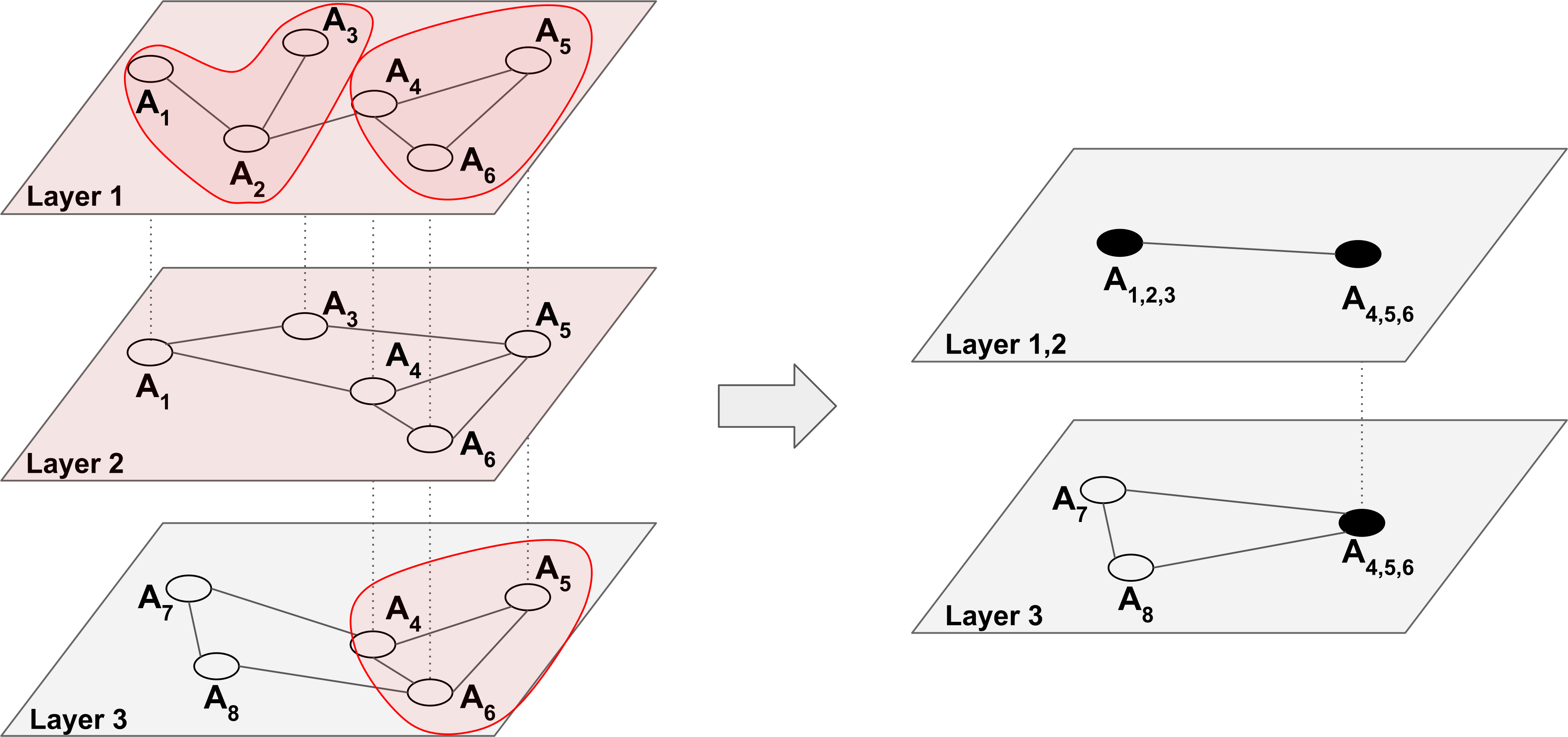}
\caption{Illustrative example of aggregation-oriented simplification: the set of actors is partitioned as $\{\{A_1, A_2, A_3\}, \{A_4, A_5, A_6\}, \{A_7\}, \{A_8\}\}$, and the set of layers is partitioned as  $\{\{L_1, L_2\}, \{L_3\}\}$ (on the left). In the simplified network (on the right), filled circles denote meta-nodes resulting from the actor aggregation.}  
\label{fig:aggregation} 
\end{figure}

\textit{Aggregation} methods 
are in principle characterized by the evaluation of expressions of type \textsf{group-by} over $\mathcal{A}$ and/or  $\mathcal{L}$, in order to obtain a multilayer network whose actors (resp. layers) are organized into a set of subsets  induced from the original multilayer network.  Note that the set of subsets of actors (resp. layers) in the simplified multilayer network is supposed to be smaller than the corresponding set   in the original network; also, this reflects on the cardinality of the sets of nodes and/or edges in the resulting simplified network.   
 An example of aggregation is shown in Figure~\ref{fig:aggregation}.

\begin{definition}[Aggregation-oriented Network Simplification]\label{def:aggregation_ns}
Given a multilayer network graph  $G = (\mathcal{A}, \mathcal{L}, V, E)$, let  $f_{\theta}$ be a simplification function with  $\theta \subseteq \Theta(\mathcal{A}, \mathcal{L}, V)$ such that $\Theta$ is a set of expressions of the form $\textsf{group}(X)$, with variables $X \in \{\mathcal{A}, \mathcal{L}, V\}$.  
The  \textit{Aggregation-oriented} network simplification problem  is to obtain a simplified network $G'=(\mathcal{A}', \mathcal{L}', V',E')$, where $G \xmapsto{f_{\theta}} G'$,     s.t. $|\mathcal{A}'| < |\mathcal{A}| \lor |\mathcal{L}'| < |\mathcal{L}| \lor
|V'| < |V| \lor |E'| < |E|$, 
where $\mathcal{A}' \subset \powerset(\mathcal{A}), \mathcal{L}' \subset \powerset(\mathcal{L})$.
\end{definition}

\begin{figure}[t!]
\centering
\includegraphics[width=1\textwidth]{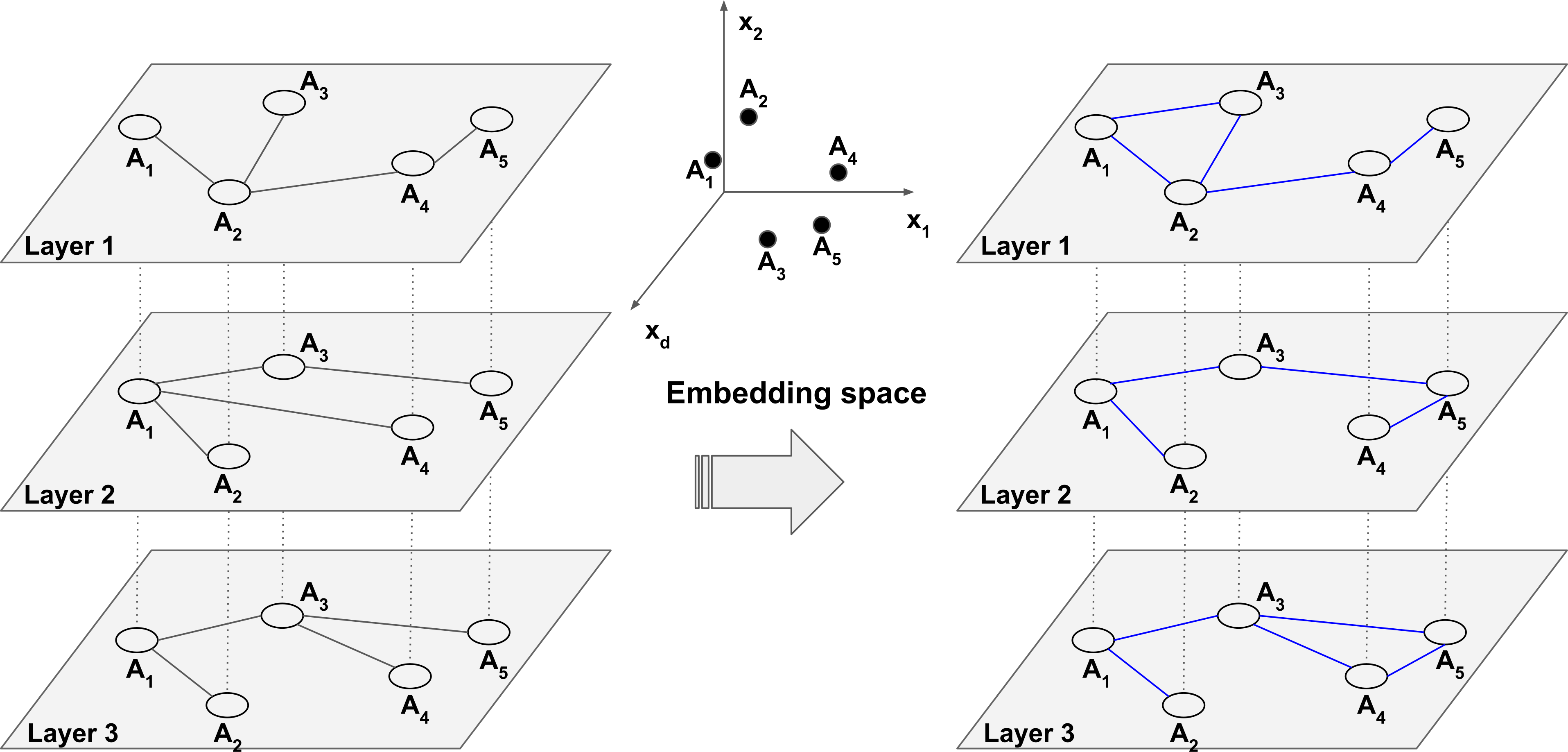}
\caption{Illustrative example of embedding-based transformation-oriented simplification: through an embedding method a new representational space is learned, in which every actor is represented by an embedding vector. In the simplified network (on the right), blue edges denote predicted edges resulting from the reconstruction process of the network through a distance measure.  
}
\label{fig:embedding} 
\end{figure}

 \begin{figure}[t!]
\centering
\includegraphics[width=0.95\textwidth]{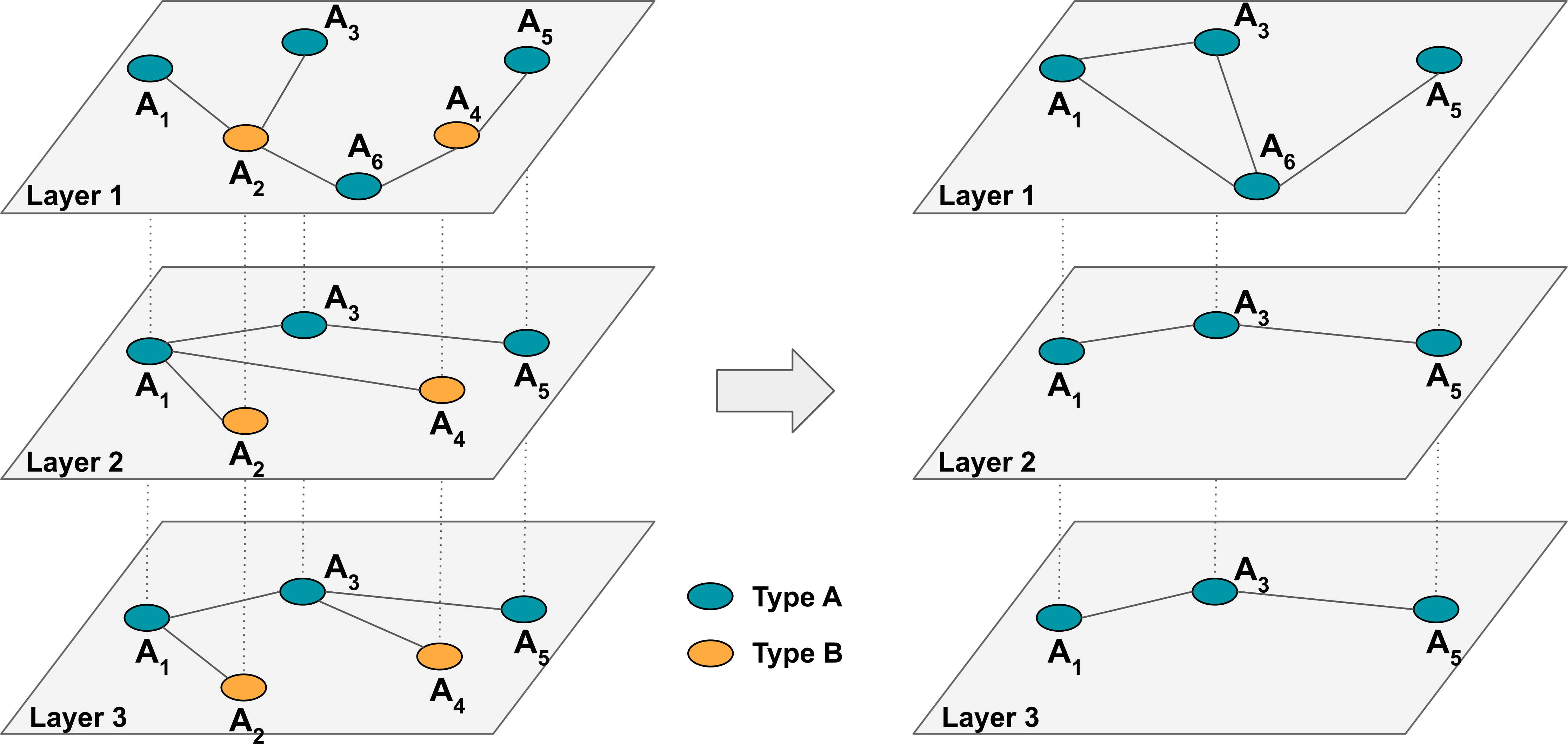}
\caption{Illustrative example of projection-based transformation-oriented simplification: given two types of actors, $A$ and $B$, in the input multilayer network (on the left), projection on type $B$ is performed (on the right). 
 \textit{(Best viewed in color version)}}
\label{fig:projection} 
\end{figure}

In the case of \textit{transformation}, we distinguish two subcategories: \textit{embedding} and \textit{projection}. 
The embedding-based transformation is characterized by a two-stage process. The first stage is to learn a low-dimensional vectorial representation (i.e., embedding) for actors and/or nodes of the input multilayer network. By exploiting the learned embeddings,  an edge is defined in the simplified network whenever the distance between any two actors' (resp. nodes') embeddings does not exceed  a predetermined threshold $\epsilon$. 
 Figure~\ref{fig:embedding} shows an example of embedding-based transformation.

\begin{definition}[Embedding-based Transformation-oriented   Network  Simplification]\label{def:transformation_ns_embed}
Given a multilayer network graph  $G = (\mathcal{A}, \mathcal{L}, V, E)$, let  $f_{\theta}$ be a simplification function with  $\theta \subseteq \Theta(\mathcal{A}, V)$ such that $\Theta$ is a set of expressions of the form 
$\textsf{dist}(X_1,X_2,\epsilon)$, with variables $X_1, X_2 \in \kappa$ such that $\kappa = g(\mathcal{A})$ (resp. $\kappa = g(V)$),   numeric constant $\epsilon \in [0,1]$, with  $g: \mathcal{A} \mapsto \mathbb{R}^d$  (resp. $g: V \mapsto \mathbb{R}^d$) with numeric constant $d>1$. The semantics of $\textsf{dist}(X_1,X_2,\epsilon)$ is that ``if the distance between $X_1$ and $X_2$ is not above $\epsilon$ then an edge is created between the actors (resp. nodes) corresponding to $X_1$ and $X_2$''. 
 The  \textit{Embedding-based Transformation-oriented} network simplification problem is to obtain a simplified network $G'=(\mathcal{A}', \mathcal{L}', V',E')$, where $G \xmapsto{f_{\theta}} G'$,    s.t. $\mathcal{A}' \subset   \mathcal{A}  \lor  \mathcal{L}' \subset  \mathcal{L}  \lor V' \subset  V$, and $E' \subseteq V' \times V'$.  
\end{definition}

The projection-based transformation aims at simplifying the multilayer network structure according to side information relating to the \textit{type} of actors (resp. nodes) present in the network. The simplification process here consists in replacing every actor (resp. node) of a given type in $\mathcal{T}$, say $\tau$, with a new edge connecting each pair of its neighboring nodes of type other than $\tau$. When removing actors (resp. nodes)  of type $\tau$, every edge connecting the removed nodes are also removed from the resulting network. Note that the   simplified network will be characterized by a subset of actors (resp. nodes) of the original network, a partially new set of edges, and an unchanged set of layers. Figure~\ref{fig:projection} shows an example of projection-based transformation.

  \begin{definition} [Projection-based Transformation-oriented Network Simplification]\label{def:transformation_prj}
Given a multilayer network graph  $G = (\mathcal{A}, \mathcal{L}, V, E)$, let  $f_{\theta}$ be a simplification function with  $\theta \subseteq \Theta(\mathcal{A}, V)$ such that $\Theta$ is a set of expressions of the form $\textsf{prj}(X)$, with variables $X \in g(\mathcal{A})$ (resp. $X \in V$), and   $g: \mathcal{A} \mapsto \mathcal{T}$ (resp. $g: V \mapsto \mathcal{T}$), with $\mathcal{T}$ set of string constants. 
 The semantics of $\textsf{prj}(X)$ is that ``every actor (resp.  node) of type $X$ is removed and an edge is created for each pair of its neighboring nodes of type other than $X$.''
 The  \textit{Projection-based Transformation-oriented} network simplification problem is to obtain a simplified network $G'=(\mathcal{A}', \mathcal{L} , V',E')$, where $G \xmapsto{f_{\theta}} G'$,    s.t. $\mathcal{A}' \subset   \mathcal{A}  \wedge  
 |E'| < |E|$, where $E' = E^{(1)} \cup E^{(2)}$, with $E^{(1)} \subset E, E^{(2)} \subseteq V' \times V'$, $V' \subset V$.
 \end{definition}

\section{Tidying Up Network Simplification Literature}
\label{sec:literature}

 In this section we elaborate on  each of the previously presented network simplification categories  and relating methods existing in the literature.  
 
\begin{figure}[t!]
\scalebox{0.58}{
\begin{forest}
for tree={%
    edge path={\noexpand\path[\forestoption{edge}] (\forestOve{\forestove{@parent}}{name}.parent anchor) -- +(0,-12pt)-| (\forestove{name}.child anchor)\forestoption{edge label};}
}
[
\textbf{Network Simplification}, calign=child,calign child=2
[\textsc{Selection}
[Filtering]
[Sampling]
]
[\textsc{Aggregation}
[Flattening]
[Community detec.]
[Coarsening]
[Positional equiv.]
[Compression]
]
[\textsc{Transformation} 
[Projection
]
[Embedding]
]
]
\end{forest}
}
\caption{Hierarchy of network simplification  approaches}\label{fig:netman}
\end{figure}
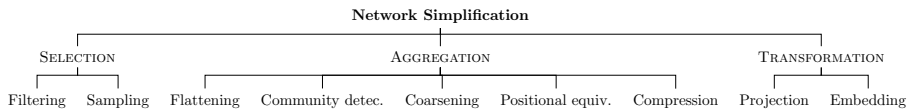

 Figure~\ref{fig:netman} shows our hierarchy of categories and subcategories of simplification techniques.   
  Moreover, as a guide to our discussion,  Table~\ref{tab:cat}  reports on main characteristics of the approaches developed for network simplification, organized according to the above provided categorization.  
For each method, the table shows: the type of information which is primarily used to drive  the simplification task; the network aspects affected by the process, where  superscript $+$ (resp. $-$) indicates that  the size of a set will increase (resp. decrease) at the end of the simplification task; 
whether the simplification scheme is deterministic or probabilistic (regardless of possible requirements in terms of meta-structures to compute, such as node groupings or embeddings);   whether the method is reversible or not, i.e., whether the original network can be fully or only partially reconstructed from the simplified one;  main references for   single-layer networks; and main references for   multilayer networks. Note that for the two last columns, we use symbol $-$ to denote the method does not make sense for  single-layer, resp. multilayer, networks, and symbol \ding{55}  to denote that no method has been developed yet. A \checkmark symbol has been used  
to denote that a well known family of techniques exists, but it is so vast that a reference list would not be exhaustive or significant.
Finally, the notation $E_\kappa$ reported for embedding-based methods represents an edge set derived from an embedding space $k$ over $\mathcal{A}$ or $V$ (cf. Definition~\ref{def:transformation_ns_embed}).

 \begin{landscape}
\begin{table} 
    \caption{Characteristics of network simplification methods. }
    \label{tab:cat}
\centering
   \scalebox{0.75}{
 \begin{tabular}{|l|l||c|c|c|c|p{5cm}|p{5cm}|}
\hline \hline 
\textbf{category} & \textbf{method} & \rotatebox{90}{\textbf{key-enabling elements}} &  \rotatebox{90}{\textbf{affected elements}}    &  \rotatebox{90}{\textbf{determinism}} & \rotatebox{90}{\textbf{reversibility}} & \textbf{single-layer native} & \textbf{multilayer native}
\\ 
\hline 
\multirow{4}{*}{Selection}   &  centrality-based filtering (cf. Sec.~\ref{par:centrality_filt}, p.~\pageref{par:centrality_filt}) & $V$  & $V^-$,$E^-$   & \Y & \N   & \checkmark & \cite{BrodkaSKM11,Sole14,Chakraborty16,Battiston16,DeDomenico14,TavassoliZ16,GalimbertiBG17,sole2016random,li2018evidential,wang2018new,MoormanCTBB18,BasarasIKT19,DingL18_topological,DingL18} \\
	&     node-layer relevance filtering (cf. Sec.~\ref{par:nodelayer_filt}, p.~\pageref{par:nodelayer_filt})    & $V$ & $V^-$,$E^-$ & \Y & \N   & -- & \cite{Rossi2015} \\  
	&	 model-based filtering  (cf. Sec.~\ref{par:model_filt}, p.~\pageref{par:model_filt}) & $V,E$  & $V^-$,$E^-$  &  \textbf{Y}/\textbf{N} & \N    & \cite{Serrano2009,Radicchi2011,Mastrandrea2014,Dianati2016,Squartini2015,Gemmetto2017,Casiraghi2017} & \cite{MandaglioAT18} \\ 
    	&	 sampling  (cf. Sec.~\ref{sec:sampling}, p.~\pageref{sec:sampling}) & $V$,$E$  & $V^-$,$E^-$ & \N & \N    & \cite{LeeKJ2006,Leskovec2005,Leskovec2006} & \cite{Gjoka2011,KhadangiBS16} \\  
    \hline
\multirow{7}{*}{Aggregation}		& community-detection-based  (cf. Sec.~\ref{sec:gpart}, p.~\pageref{sec:gpart})  & $V,E$ & $V^-$,$E^-$  & \textbf{Y}/\textbf{N}  & \N   & \cite{Newman2004,Clauset2004} & \cite{pmm,KimL15,Loe15,Kivela+14,PhysRevX.5.011027,AfsarmaneshM16,TangWL12,KunchevaM15,PapalexakisAI13,BerlingerioPC13,JeubMMP15,InterdonatoTISP17} \\ 
	&	 multilevel partitioning  (cf. Sec.~\ref{sec:coarsening}, p.~\pageref{sec:coarsening})   & $E$ & $V^-$,$E^-$  & \textbf{Y}/\textbf{N} & \Y    & \cite{Hendrickson1995,Gupta1997,Holtgrewe2010,Sanders2011,Safro2015,Glantz2016,Osipov2010,Bui1993,Hendrickson1995,Karypis1998,LaSalleK15,Fruchterman1991} & --\\ 
	&    positional equivalence  (cf. Sec.~\ref{sec:poseq}, p.~\pageref{sec:poseq})   & $E$ & $V^-$,$E^-$  & \Y & \Y   & \cite{Holland1981,Faust1988,Everett1991,lorrian1971structural:methods,LUCZKOVICH2003,Wasserman1994,ryan2015roles:sna,peixoto2017bayesian,peixoto2018nonparametric,Henderson2012,Doreian2004} & \cite{ZIBERNA201446,Stanley2016,paul2016consistent,pamfil2018relating}   \\  
	&    flattening   (cf. Sec.~\ref{sec:flat}, p.~\pageref{sec:flat})  & $\mathcal{L}$ & $\tau_E^-$,$V^-$,$E^-$  & \Y & \N    & --  & \cite{Dickison2016,BerlingerioCG11} \\   
	&   layer aggregation  (cf. Sec.~\ref{sec:flat}, p.~\pageref{sec:flat})  & $\mathcal{L}$ & $\tau_E^-$,$E^-$,$V^-$  & \Y  & \N    &  -- & \cite{DeDomenico2015} \\  
 	&  graph compression  (cf. Sec.~\ref{sec:compr}, p.~\pageref{sec:compr})   & $V$,$E$ & $V^-$,$E^-$  & \Y & \Y  & \cite{Navlakha2008,Toivonen2011,Lim2014,Adler2001,Feder1995,Fan2012,Maneth2015ASO,LiuSDK18,Khan2016,Nourbakhsh2014,MANETH201819,Shah2015} & \ding{55}   \\  
	& coarsening   (cf. Sec.~\ref{sec:coarsening}, p.~\pageref{sec:coarsening})  & $V,E$ & $V^-,E^-$ & \textbf{Y}/\textbf{N} & \Y   & \cite{Geisberger2012,Basak2016,Buluc2016,Walshaw2003,Hachul2005,hu2005efficient,Martin2011} & \ding{55}    \\  
    \hline     
\multirow{2}{*}{Transformation} 
     &	 projection-based (cf. Sec.~\ref{sec:proj}, p.~\pageref{sec:proj})   & $\mathcal{A}|V$, $\mathcal{T}$ &  $V^-$, $E^+$  & \Y & \N   &  \checkmark & \cite{Seierstad2011,Padron2011,Newman2001,Opsahl2013} \\ 
    & embedding-based (cf. Sec.~\ref{subsec:embedding}, p.~\pageref{subsec:embedding})  & $\mathcal{A}|V$, $\kappa$  & $V$,  $E_\kappa$ &  \textbf{Y}/\textbf{N}& \N    & \cite{deepwalk,node2vec,line,graphsage,gcn,sae,dngr,sdne} & \cite{PMNE,MNE,MELL} \\  \hline 
    \end{tabular}  
  } 
\end{table}
 
 \end{landscape}

 \begin{figure}[t!]
\centering
\subfigure[]{\includegraphics[width=0.45\textwidth]{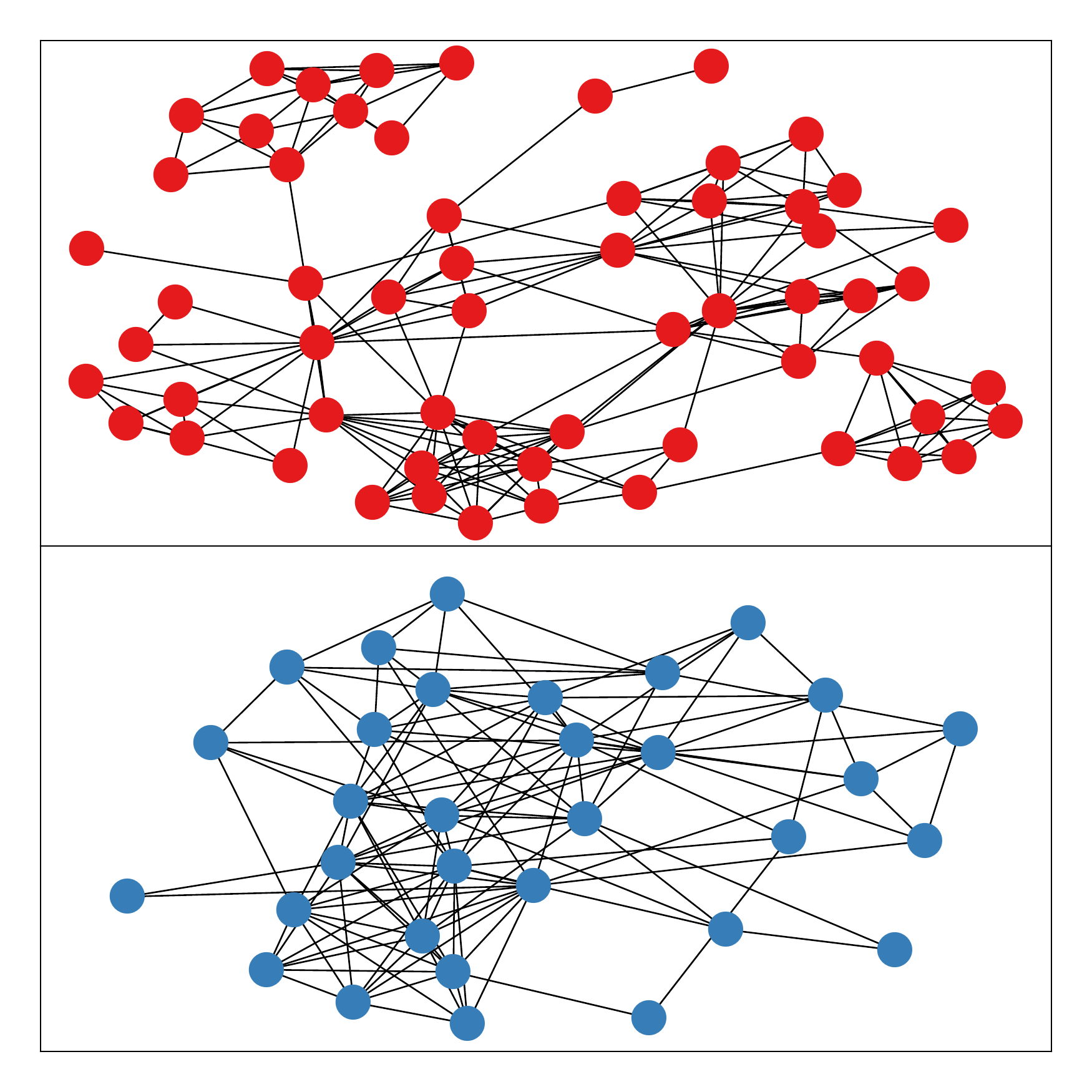}} 
\subfigure[]{\includegraphics[width=0.45\textwidth]{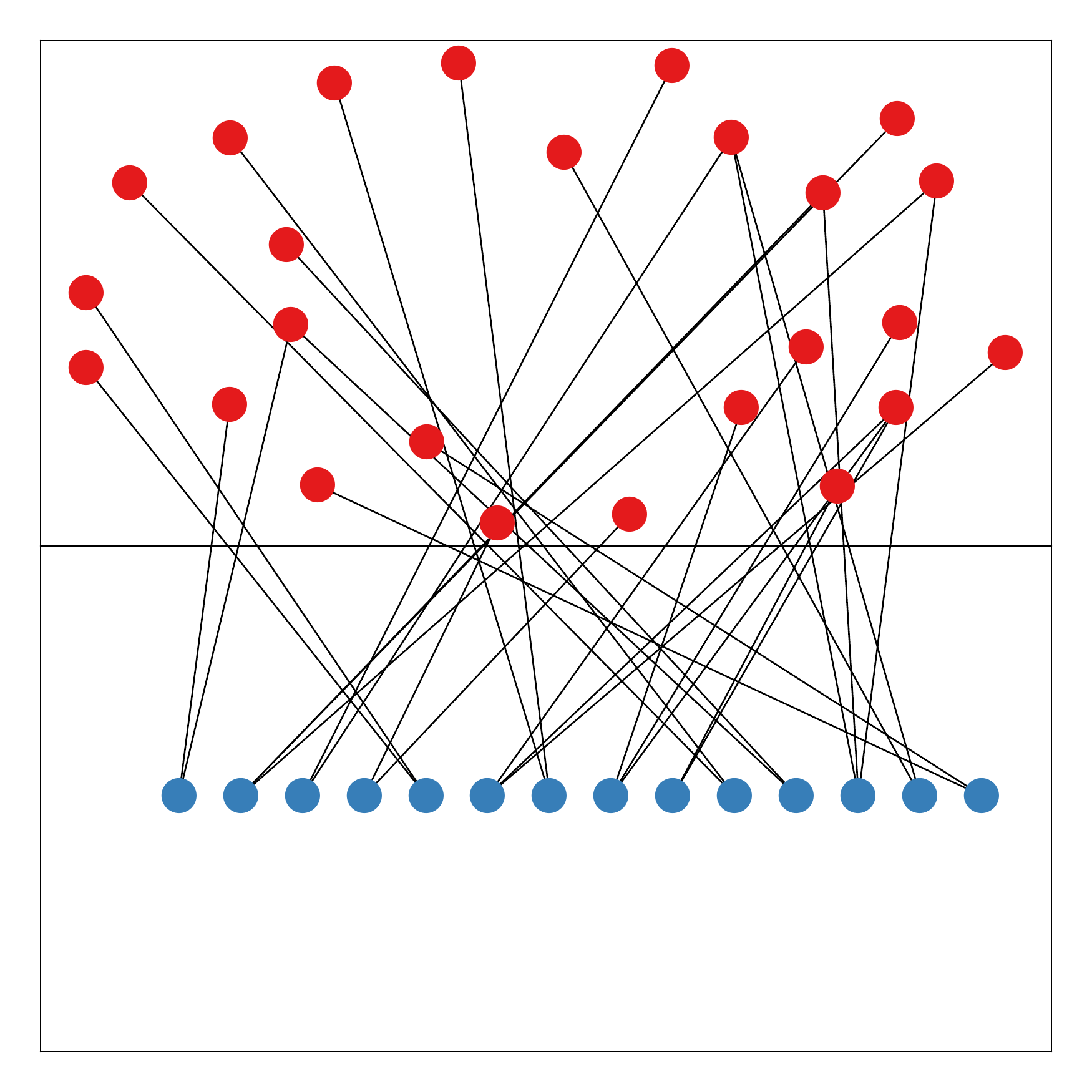}} 
\caption{Two portions of the AUCS network used to exemplify simplification methods. (a) Two multiplex layers with ``lunch'' (top) and ``Facebook'' (bottom) relationships between the same set of employees. (b) Two heterogeneous layers with employees (top) and research papers (bottom), with edges indicating who authored which paper.}
\label{fig:example_data} 
\end{figure}

In the following we will use a real multilayer network about a university department \cite{Rossi2015} to exemplify different simplification methods. In this network different layers are used to represent either different types of edges between the same actors, e.g., having lunch together and being friends on Facebook, or edges between different types of entities, in particular authorship relationships between employees and research papers. Figure~\ref{fig:example_data} shows two combinations of layers that will be used as running examples. 

\subsection{Selection based techniques}

We use the broad term   \textit{selection} to refer  to  techniques that aim to reduce the size of the input network  by selecting a subset of nodes, edges, or layers, according to  specific features involving the interested  entities of the network. 
Selection techniques can be grouped into two main categories: filtering and sampling. The former includes methods that exploit selective information requirements, such as methods that employ   information given by centrality measures as a selection criterion, and approaches that leverage edge statistical significance information to filter out noisy links. The latter contains   approaches belonging to random-access and exploration-based categories.

\subsubsection{Filtering}
\label{sec:filtering}
  
Filtering approaches resemble the problem of \textit{dimensionality reduction} in relational data, in that they  involve  the identification and  removal of a (possibly large) number of edges or nodes that are supposed to  prevent the discovery of patterns due to redundancy, irrelevance, missing values, noise or other related aspects.  
Filtering in graphs is also commonly referred to by using    different terms  such as pruning, thresholding, and sparsification. 
 In particular, thresholding refers to filtering out those edges/nodes whose value for some selected property exceeds a predefined threshold, while sparsification typically aims to approximate the input graph with a sparser one, guaranteeing that some properties  
  of the original graph are preserved within a chosen degree of  tolerance. 
  For the sake of simplicity, in the rest of this section we will use filtering as   unifying term, unless otherwise specified.

\paragraph{Centrality-based filtering}
\label{par:centrality_filt} 

The basic idea underlying centrality-based filtering is   to apply one of the aforementioned approaches to rank the nodes by centrality scores and, given a predetermined threshold, select those nodes in a graph that satisfy the constraint based on the threshold. 
The term \textit{centrality} in networks commonly refers to importance or prominence of a node in a network, i.e., the status of being located in strategic locations within the network. There is no unique definition of centrality, as for instance one may postulate that a node is important if it is involved in many direct interactions (i.e., \textit{degree} centrality), or if it connects two large components acting as a bridge (i.e., \textit{betweenness} centrality), or if it allows for quick transfer of the information also by accounting for indirect paths that involve intermediaries (i.e., \textit{closeness} centrality). 

Brodka et al. propose  a definition of  degree centrality for multilayer networks in~\cite{BrodkaSKM11}. 
Sole et al.~\cite{Sole14} define  a multilayer extension of the betweenness centrality by taking into account shortest paths that include inter-layer edges. 
The same authors in~\cite{sole2016random}  extend  random walk betweenness and closeness centrality  to interconnected multilayer networks. 
A cross-layer betweenness centrality is proposed in~\cite{Chakraborty16}, also including applications to multilayer community detection and message spreading tasks.
Two classes of multilayer degree-biased random walks are proposed in~\cite{Battiston16}, to analyze to what extent this kind of random walks can make the exploration of multilayer networks  more efficient. 
In~\cite{DeDomenico14}, De Domenico et al.  study   main factors   influencing the navigability of multilayer networks, using random walks over a layer-aggregated network; in general, when dealing with centrality scores produced at distinct layer networks,  the use of different aggregation and normalization techniques is shown to  strongly bias the final results~\cite{TavassoliZ16}.  
In~\cite{DingL18_topological}, Ding and Li define topologically-biased random walks on multiplex networks and derive analytical expressions for their long-term diffusion properties such as entropy rate and stationary probability distribution. They found  that inter-layer coupling strength, edge overlapping, the sign and presence of inter-layer degree-degree correlations and the number of layers  capture the extent to which the diffusions on a multiplex network are efficiently explored by a biased walk. 
The same authors define topologically biased multiplex PageRank in \cite{DingL18}. 
Perna et al. in~\cite{PernaIT18}  introduce the Alternate Lurker-Contributor Ranking method (mlALCR). By solving two mutually dependent systems of equations, mlALCR is able to identify users characterized by alternate behaviors (i.e., information-producer vs. information-consumer) across the layers of a   multilayer  network. 
 Using a fourth-order tensor to represent multilayer networks, Wang and Zou~\cite{wang2018new}   propose   the Singular Vector of Tensor (SVT) centrality,  which is used to quantitatively evaluate the importance of nodes connected by different types of links in multilayer networks, through the computation of   hub and authority scores of nodes and layers in multilayer networked systems.
 In~\cite{BasarasIKT19}, Basaras et al.  study how to identify influential information spreaders in multilayer   networks, and they propose a family of measures for describing the strategic position of a node within a multilayer network, based on the connectivity of the node with respect to nodes belonging to the same layer as well as to the rest of the layers.

A related approach, which is also commonly adopted in simple graphs, is to induce an organization of the nodes into substructures with desired structural characteristics based on some notion of centrality. Within this view, an exemplary method is    
the \textit{k-core} decomposition, which consists in   finding cohesive subgraphs based on node degree \cite{Seidman1983}. A \textit{k-core} is a maximal subgraph in which all nodes have degree at least $k$.   The problem of core decomposition of a multilayer network is studied in \cite{GalimbertiBG17}. Given  a multilayer network $G = (\mathcal{A}, \mathcal{L}, V, E)$ and an $|\mathcal{L}|$-dimensional integer vector $\mathbf{k} = [k_l]_{l \in \mathcal{L}}$,  the multilayer $k$-core of $G$ is the maximal subgraph whose nodes have at least degree $k_l$ in that subgraph, for all layers $l$. Vector $\mathbf{k}$ is the coreness vector of the $k$-core. The set of all non-empty and distinct multilayer cores constitutes the multilayer core decomposition of $G$.
Nevertheless, the authors observe that with this definition the number of multilayer cores can be exponential in the number of layers, i.e., the cores are not nested into each other like in the single-layer case, rather they form a core lattice defining a relation of partial containment. As a solution, the authors propose three algorithms based on different pruning rules and visiting strategies of the lattice. 
In~\cite{MoormanCTBB18}, Moorman et al.  propose effective filtering methods for finding subgraphs isomorphic to an input template graph inside  a large multiplex network by reducing the search space based on local statistics.

 \begin{figure}[t!]
\centering
\subfigure[]{\includegraphics[width=0.45\textwidth]{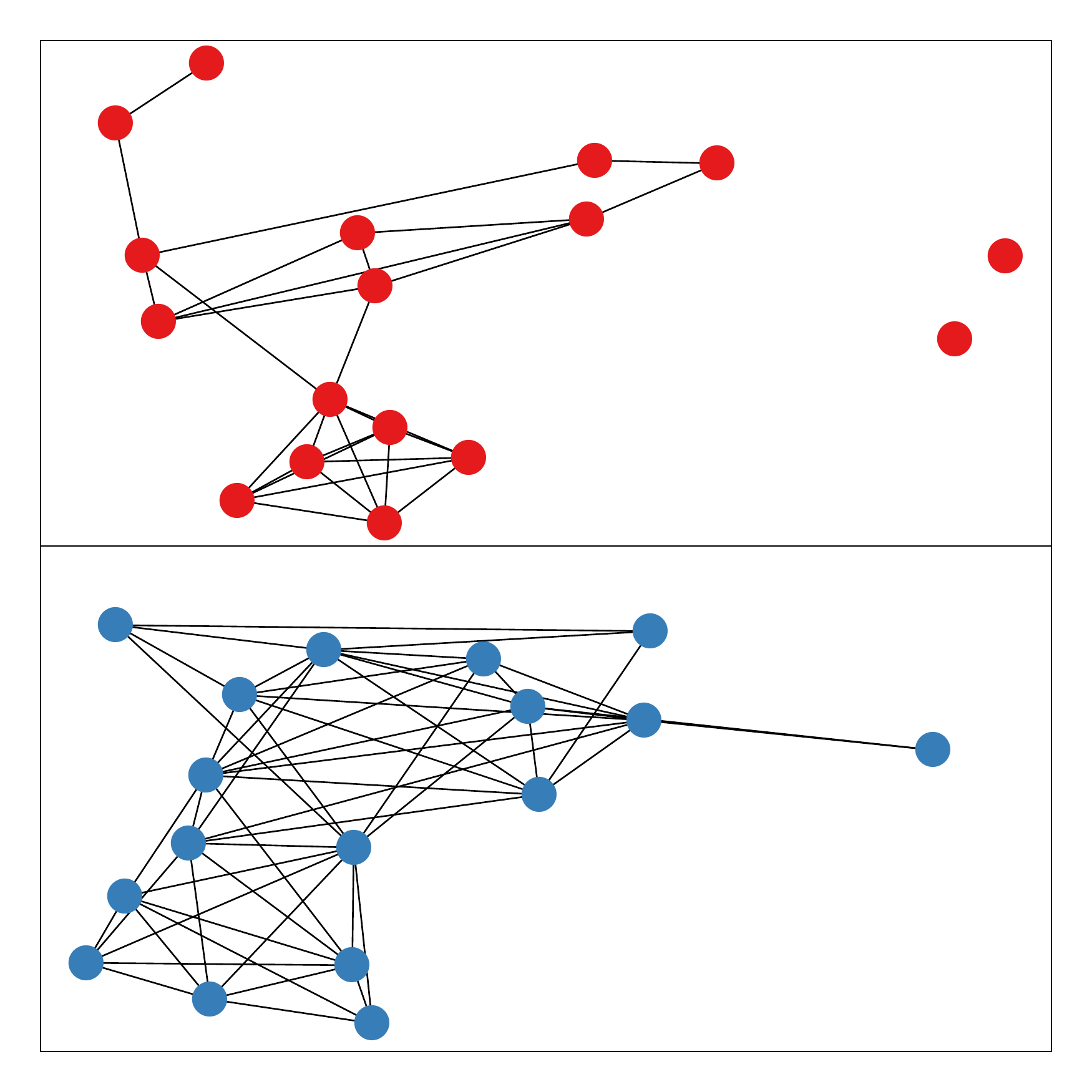}} 
\subfigure[]{\includegraphics[width=0.45\textwidth]{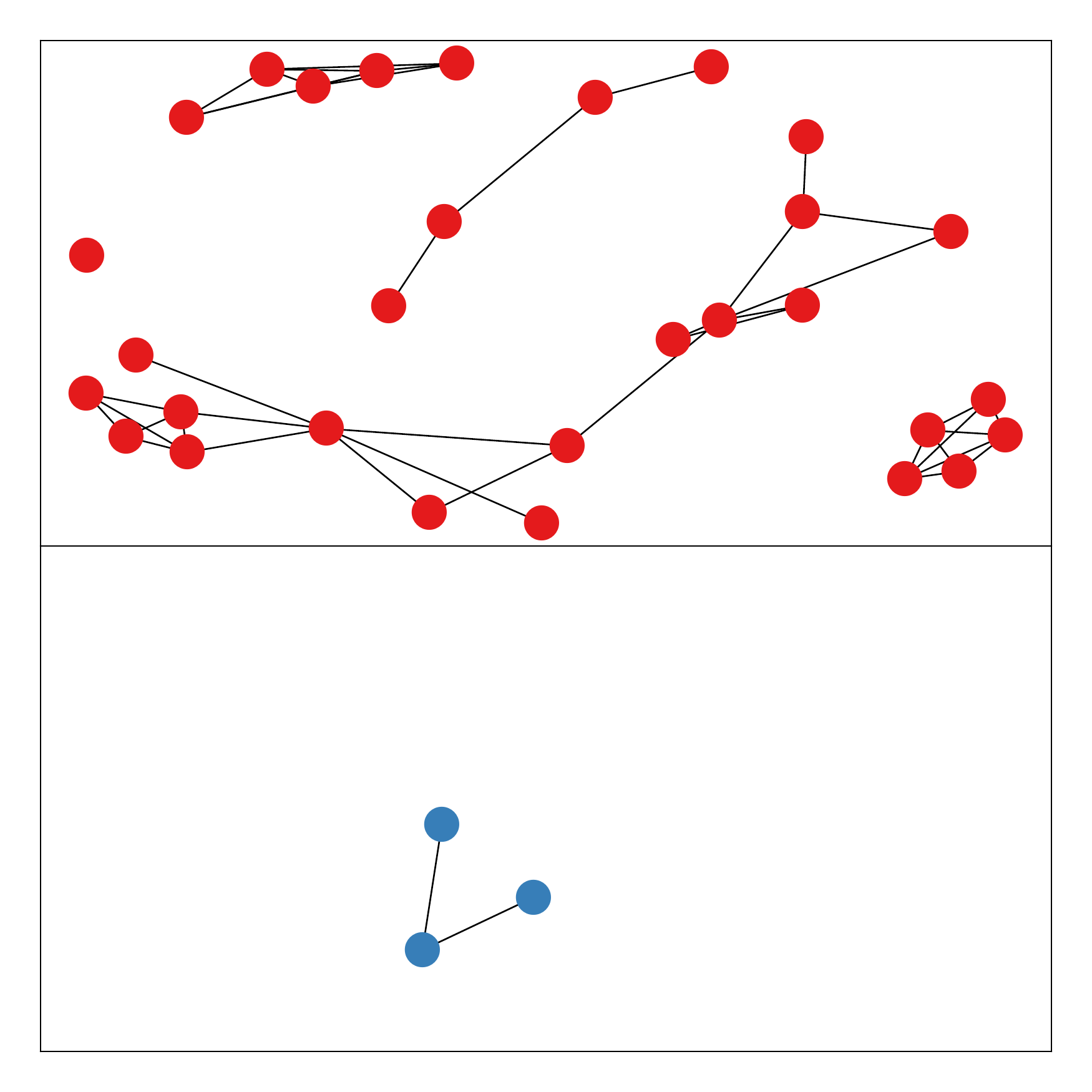}} 
\caption{The ``lunch'' and ``Facebook'' layers in the AUCS networks filtered according to (a) neighborhood, in particular, preserving only nodes with at least six neighbors in both layers, and (b) exclusive layer relevance greater than .6, indicating that at least $60\%$ of the actor's neighbors on that layer are only present on that layer.}
\label{fig:example_filtering} 
\end{figure}

Examples of different types of multilayer node filtering are shown in Figure~\ref{fig:example_filtering}. In Figure~\ref{fig:example_filtering}(a) a node is kept on a layer only if the corresponding actor has a sufficient degree centrality in all layers. Notice that, as with simple networks, this approach may not preserve the degree centrality rank of the nodes: a node adjacent to a large number of nodes, none of which passes the filtering threshold, will be disconnected on the filtered network despite its high original degree. In Figure~\ref{fig:example_filtering}(b) only nodes having most of their neighbors exclusively present on a layer are preserved. From this simplification we can see how Facebook connections are in most cases replications of offline connections, and how some clusters exist in the lunch layer that would disappear if these connections had to rely on Facebook friendships.

\paragraph{Node-layer relevance-based filtering}
\label{par:nodelayer_filt}  
Rossi and Magnani~\cite{Rossi2015} define a local simplification approach in the context of a discussion about effective visualization strategies. 
Two variants of the approach are proposed, based on the concepts of \textit{relevance} and \textit{exclusive relevance} introduced in~\cite{BerlingerioCGMP13}. Given a node-layer pair, \textit{relevance} measures the fraction of neighbors a node has in a layer, with respect to its total number of neighbors; similarly, \textit{exclusive relevance} measure the fraction of \textit{exclusive} neighbors with respect to a layer, i.e., neighbors a node just has in that layer.
The simplification process is performed as a threshold-based filtering of the edges in the network: given a threshold, for each layer, only edges between nodes with a relevance value above the threshold in that layer are kept in the network. 
Compared to other simplification approaches for multilayer networks, this approach has the advantage of preserving the full information about the original layers.  
However to the best of our knowledge this approach has not been tested on real applications, therefore there is still limited evidence regarding its effectiveness.

\paragraph{Model-based filtering}  
\label{par:model_filt}
A relatively recent corpus of study addresses the task of filtering out ``noisy'' edges from complex networks based on \textit{generative null models}. The general idea is to define a null model based on  edge distributions, use it to compute a $p$-value for every edge (i.e., to determine the   statistical significance of properties  assigned  to   edges from a given distribution),  and finally filter out all edges having $p$-value above a chosen significance level, thus keeping all edges that are least likely to have occurred due to random chance.

Methods following the above general approach have been mainly conceived to deal with weighted networks, so that both the node degree and node strength (i.e., the sum of the weights of all incident edges) are used to generate a model that defines a random ensemble of graphs resembling the observed network.   
One of the earliest methods is the \textit{disparity filter}~\cite{Serrano2009}, which evaluates the strength and degree of each node locally. The null hypothesis is that the strength of a node is redistributed uniformly at random over the node's incident edges. The disparity filter hence introduces some bias in that the strength of neighbors of a node are discarded.
By contrast, a global null model is defined with the \textit{GloSS filter}~\cite{Radicchi2011},  
    as it preserves the whole distribution of edge weights within a single graph. The null model is in fact  a graph with the same topological structure of the original network and with edge weights randomly drawn from the empirical weight distribution. Since all edges have the same probability of being assigned a given weight, the statistical test is the same for every edge, and hence this reduces to setting a global threshold (depending on a chosen significance level) for pruning. 
Unlike disparity and GloSS, the null model proposed by Dianati~\cite{Dianati2016} 
is  maximum-entropy based and hence unbiased. Upon it, two filters are defined:  the  marginal likelihood filter, which is a linear-cost method that assigns  a significance score to each edge based on the marginal distribution of edge weights, and  the global likelihood filter, which is an ensemble approach that  accounts for the  correlations among edges. Both filters consider   the strength of nodes, but not the degrees.  
Recently, Gemmetto et al.~\cite{Gemmetto2017}  
proposed a maximum-entropy filter that guarantees that only irreducible edges are kept, i.e.,    the filtered network will retain only the edges that cannot be inferred from local information.   The general goal is to unveil  the   backbone of non-redundant structures in a complex network. The proposed filter  employs a null model based on the canonical maximum-entropy ensemble of weighted networks having the same degree and strength distribution as the real network~\cite{Mastrandrea2014}, which allows to overcome redundancy issues that arise in the aforementioned filters. 
 On a similar research direction is the unbiased method proposed in~\cite{Squartini2015}, which combines   an exact maximum-likelihood approach with an efficient computational
sampling scheme to sample ensembles of various types of networks (i.e., 
directed, undirected, weighted, binary) with many possible constraints (degree sequence, strength sequence, reciprocity structure, mixed binary and weighted properties, etc). 
 The generalized hypergeometric ensembles (gHypE) framework proposed in~\cite{Casiraghi2017} focuses instead on inferring significant links in relational data, by providing  an analytically tractable statistical model of directed
and undirected multi-edge graphs. It  can also account for known factors that influence the occurrence of interactions, such as known group structures, similarities between elements, or other forms of biases.

Some of the above generative models for graph pruning have been recently used in the context of \textit{consensus community detection} in multilayer networks~\cite{TagarelliAG17,MandaglioAT18}. Essentially, a generative null model is evaluated on a \textit{weighted graph of co-associations} (or co-occurrences): given   an input multilayer network and an ensemble of layer-specific community structures defined over it, a weighted co-association graph is an undirected graph whose nodes correspond to the actors/entities in the multilayer network and the strength of an edge corresponds to the fraction of communities that two entities share  in the ensemble community structures. The key idea adopted in~\cite{MandaglioAT18} is that a relatively low value of co-association might be retained as meaningful provided that it refers to node relations that make sense only for certain layers; by contrast,  a relatively high value of co-association could be discarded if it corresponds to the linkage of nodes that have high degree and co-occur in the same community in many layers, and hence  the co-association could be considered as superfluous in terms of community structure.

\subsubsection{Sampling} \label{sec:sampling}
\label{sec:sampling} 
When dealing with the analysis of real-world complex networks, it is likely to incur in scalability issues. In these cases, a popular solution is to perform the analysis on a smaller sample of the network.
With the term \textit{network sampling} we will refer to all the techniques which aim at obtaining a (relatively small) representative sample from a network, i.e., a subset of nodes and edges partially preserving structural characteristics of the original network.
Sampling can also be thought as a valid alternative to synthetic graph generation \cite{Krishnamurthy2005}, i.e., instead of growing a synthetic graph with a set of desired properties, in some cases it can be preferable to shrink the original graph to a smaller size maintaining the main structural properties.
 
It should be noted that while network sampling and \textit{graph compression} (cf. Section~\ref{sec:compr}) are often used to refer to similar methods in literature, in this work we will consider as \textit{network sampling} all the techniques based on stochastic/random processes (i.e., non-deterministic techniques), while techniques which include deterministic procedures allowing   reversibility or recovery-rate of the process will be considered in the   \textit{graph compression} category.  

While the most common scenario is that of a scale-down goal (i.e., obtaining a sample showing similar properties as compared to the original graph), the effectiveness of a sampling process may also be evaluated considering a back-in-time goal, i.e., when the aim is to obtain a sample similar to what the original graph looked like when it was the size of the sample~\cite{Leskovec2006}.

Sampling techniques can be categorized based on the way they access the graph during the sampling process, i.e., random access based techniques or exploration based techniques.
While the former envisages a local approach, the latter includes all the approaches which take into account broader information about the graph structure. In the following, we provide an overview of  main sampling approaches for each category.

\paragraph{Random access techniques} 
An intuitive way to sample a graph is to perform a random selection process, i.e., fixed a size $n$ for the sample, uniformly select $n$ nodes (or $n$ edges) at random and build their induced subgraph~\cite{Leskovec2006}. Nevertheless, sampled graphs obtained using this technique do not retain basic graph properties, e.g., random subnets sampled from scale-free networks are not themselves scale-free~\cite{Stumpf2005}. 
Non-uniform selection criteria can also be used, e.g., nodes may be selected based on their degree or PageRank score.
Hybrid strategies have also been proposed, based on random selection of both nodes and edges.
In the \textit{random node-edge sampling}~\cite{Leskovec2006}, at each step a node is randomly selected, and then one of its neighbors is randomly selected. \textit{Random edge-node sampling}~\cite{Rafiei2005} starts by selecting a set of random edges, then an induced graph is built by adding all the neighbors of the nodes connected to the original set of random edges. 
Random access techniques generally lead to degree distribution and sparse connectivity problems~\cite{Leskovec2006}. While their applicability is guaranteed also on multilayer networks (e.g., performing global or layer-wise random processes), these drawbacks are even emphasized in this scenario, where a random selection may also create a strong population and connectivity unbalance among the different layers.

\paragraph{Exploration-based techniques}
Advanced sampling techniques have been proposed in order to overcome the problems coming from random access based approaches, introducing solutions that allow to partially maintain some characteristic of the original network, e.g., network connectivity and degree distribution.
Nevertheless, while consolidated approaches exist for simple graphs, sampling of multilayer networks can be considered an open problem, since few techniques have been proposed that are specifically conceived for these models. 

Exploration based techniques (also referred to as \textit{topology-based} or  \textit{traversal-based}) for simple networks, are generally based on the idea to first select a node uniformly at random and then explore its neighborhood. A well known example of this strategy is the snowball sampling, which adds nodes and edges using breadth-first search from a randomly selected seed node. 
While the network connectivity is maintained within the snowball, many peripheral nodes (i.e., those sampled on the last round) will be missing a large number of neighbors, causing the so-called \textit{boundary bias} phenomenon \cite{LeeKJ2006}. 
Another well known exploration strategy is the random walk based one proposed in \cite{Leskovec2006}, that works by picking a starting node uniformly at random, and then sampling the graph by simulating a random walk on it.  
Sampling techniques based on temporal graph evolution have also been defined, e.g., \textit{Forest Fire} sampling~\cite{Leskovec2005}. The idea is to randomly pick a seed node
and begin ``burning'' outgoing links and the corresponding nodes, based on two input parameters called forward and backward burning probability. The process goes on recursively starting from the endpoints of the burned links, until convergence (as the
process continues, nodes cannot be visited a second time, preventing infinite loops).
  
As regards sampling approaches for multilayer networks, an early approach has been proposed by Gjoka et al.~\cite{Gjoka2011} in the context of online social networks. 
They observe that, when dealing with online social network graphs, random walk based sampling can produce representative samples only if the social graph is fully connected. To solve this problem,  they propose to take into account multiple social interactions happening in a social network, in order to have higher chances to obtain a fully connected \textit{union} graph, i.e., a single-layer network built by taking into account all possible relation types.
To further improve upon the sampling process, they then propose a layer-wise random walk based sampling, which seems to outperform standard techniques on the obtained \textit{multigraph}.  The idea is to perform a two-stage random walk based sampling, where the first stage consists in selecting a relation type on which to walk (i.e., a layer), and the second one in enumerating  the neighbors with regards to that relation only. 
More recently, Khadangi et al.~\cite{KhadangiBS16} addressed a similar sampling context taking Facebook as case in point, by proposing a biased sampling techinque for a multilayer activity network, where the activities are regarded as multiple social interactions (e.g., like, comment, post and share). The idea is to use a reinforcement learning scheme, i.e., learning automata~\cite{Rezvanian2014}, in order to learn transition probabilities among the users, and then apply a random walk-based sampling on the activity network using the learnt probabilities. The proposed approach allows to perform a biased sampling, i.e., obtain a sample subgraph consisting of suitable nodes according to application-based measures, such as fame, spam, politeness, trust, closeness, and time spent on the social network.

\begin{figure}[t!]
\centering
\subfigure[]{\includegraphics[width=0.45\textwidth]{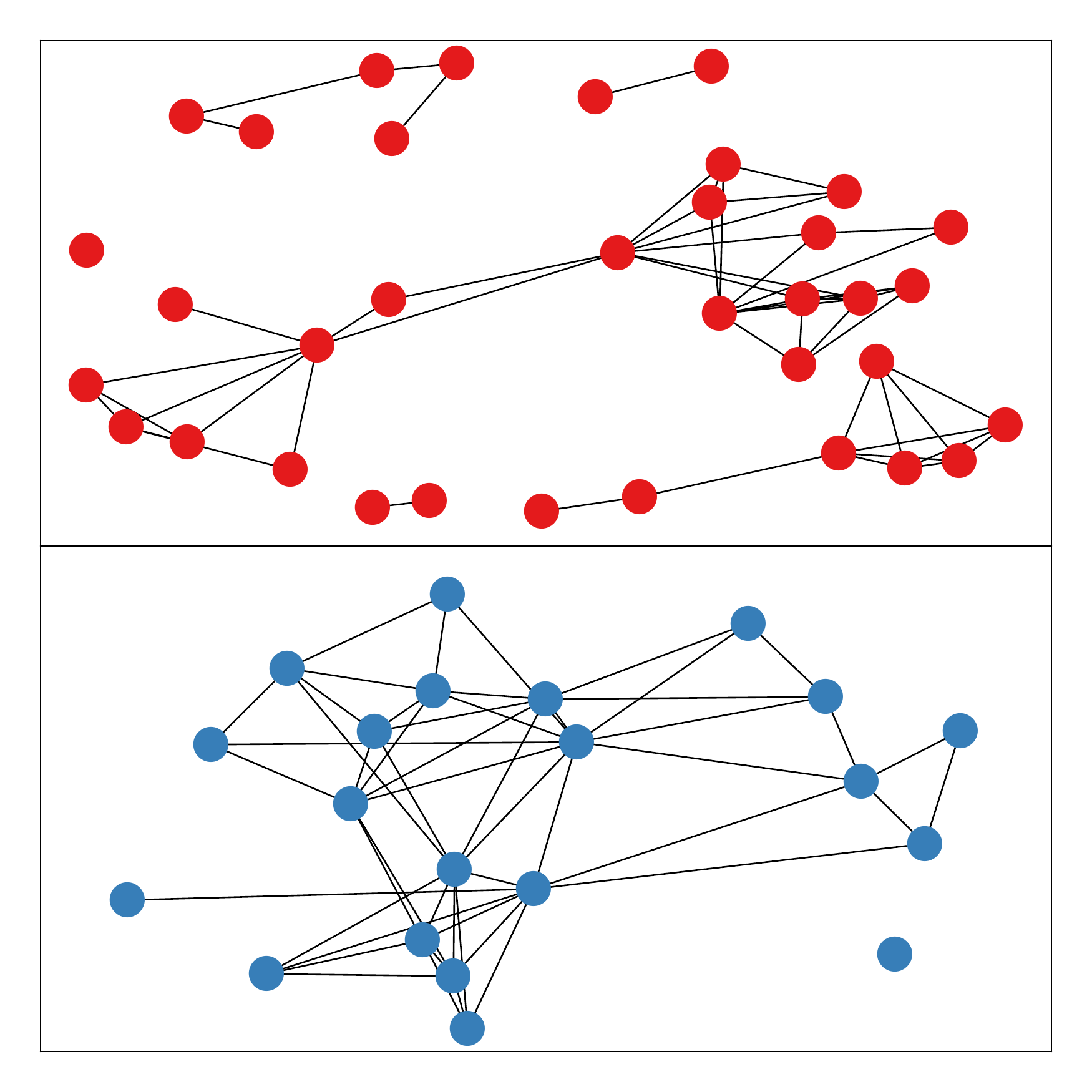}} 
\subfigure[]{\includegraphics[width=0.45\textwidth]{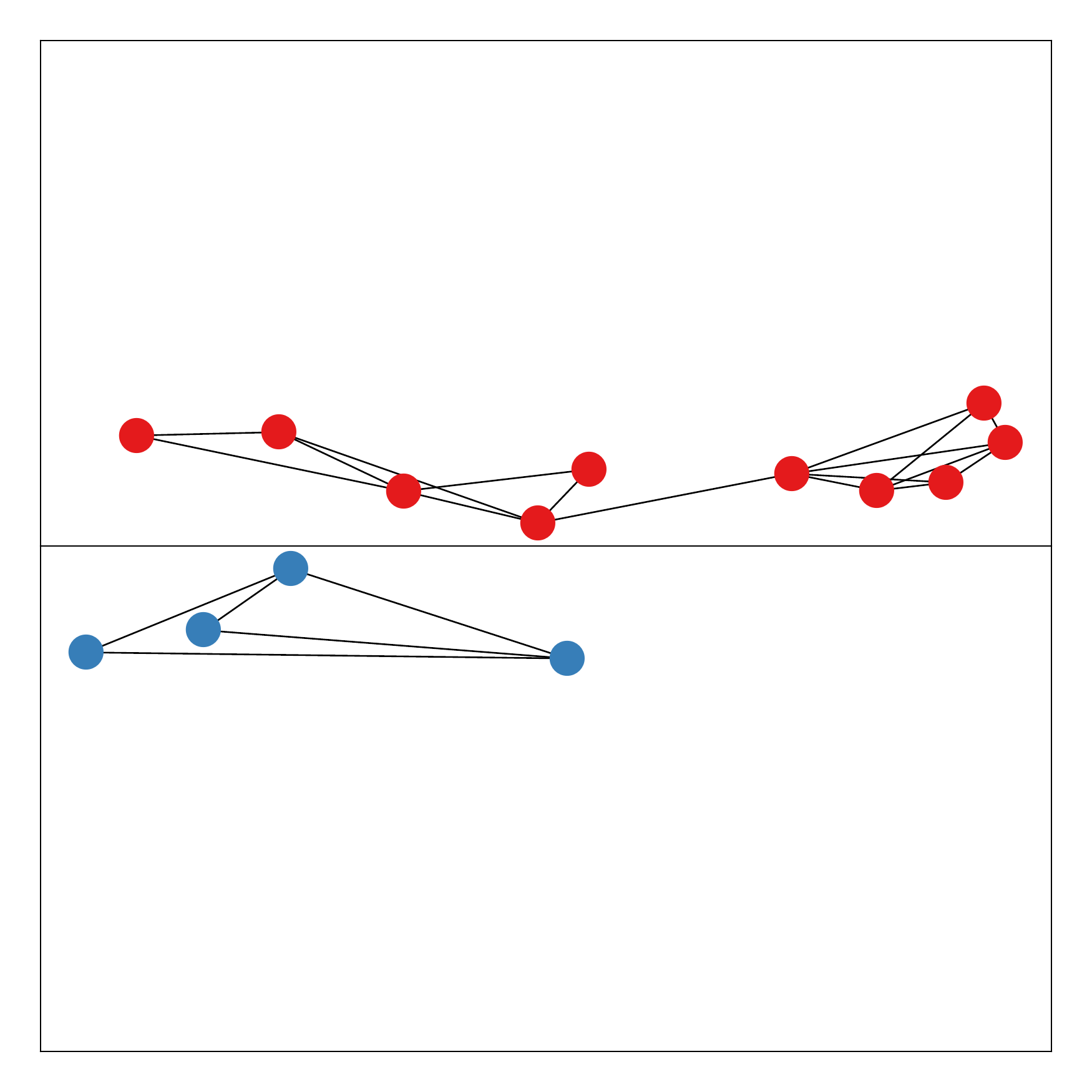}} 
\caption{The  ``lunch'' and ``Facebook'' layers in the AUCS networks sampled according to (a) random node choice and (b) snowball sampling.}
\label{fig:example_sampling} 
\end{figure}

Examples of random and snow-ball sampling simplification are shown in Figure~\ref{fig:example_sampling}. Notice that while the principles are the same as in simple networks, sampling in multilayer networks require additional details. For example, in Figure~\ref{fig:example_sampling}(b) when an actor that is present on another layer is encountered, its neighbors in the other layer are also retrieved.

\subsection{Aggregation based techniques}
Aggregation refers to techniques that aim to obtain a more compact version of the input multilayer graph by employing partitional or hierarchical grouping mechanisms that involve the building blocks of a multilayer graph (i.e., nodes, edges, layers). Note this differs  from selection techniques, which rely on the selection of a subset of nodes, edges, or layers and evaluate them individually. 

We organize our presentation of aggregation techniques into five categories, namely \textit{flattening}, \textit{community detection}, \textit{coarsening}, \textit{positional equivalence}, and \textit{graph compression and summarization}.  
Flattening methods transform the input multilayer network in a   single-layer (weighted) network  by discarding the information that characterizes the individual layers of the network, thus  aggregating them into a single layer. Multilayer community detection methods allow the definition of simplification functions to map actors/nodes to cohesive groups.  
Positional equivalence methods leverage the notions of structural equivalence and interchangeability to obtain a simplified network. Graph compression and summarization approaches, which are originally conceived to deal with   large graphs, aim to reduce the size of the graph and improve the efficiency in terms of storage and   execution time. 
 Similarly to graph compression and summarization, coarsening or graph contraction aims to enable the application of computationally expensive algorithms on large graphs by building a hierarchy of successively aggregated graphs with decreasing size. 
 In the following, we will discuss in detail each of the above categories of aggregation based techniques.

\subsubsection{Flattening}
\label{sec:flat}

When dealing with multilayer networks, handling and analyzing the network can be problematic when the number of collected layers is relatively high. Moreover, applicability of standard  network analysis techniques is not guaranteed on multilayer models. 
A straightforward solution in these cases is to \textit{flatten} the network, i.e., given a multilayer network $G = (\mathcal{A}, \mathcal{L}, V, E)$, discard the information about the layers $\mathcal{L}$. 
The result of a flattening process is a single-layer network, where the edge set contains a single instance of each relation between two nodes, i.e., all the information about edge types (e.g., different types of relations happening or not between a pair of nodes) will be lost.
As defined in \cite{Dickison2016}, a  basic (unweighted) flattening of a multilayer network $G = (\mathcal{A}, \mathcal{L}, V, E)$ is a graph $(V_f,E_f)$ where $V_f = \{a | (a,l) \in V\}$ and $E_f = \{ (a_i,a_j) | \{(a_i,l_q),(a_j,l_r)\} \in E\}$.

In order to avoid a complete loss of information about edge types, basic flattening can be improved by including edge weights proportional to the number of edges between two actors in the original network, i.e., the number of layers where a certain edge existed in the original network. 
A more sophisticated way to flatten a multilayer network can be based on the use of a weighting scheme which assigns a weight $w_{q,r}$ to each pair of layers $(l_q,l_r)$, so that the resulting single-layer network can be expressed as a linear combination of the original multilayer network.
As defined in~\cite{Dickison2016}, for a multilayer network $G = (\mathcal{A}, \mathcal{L}, V, E)$, given a $|\mathcal{L}| \times |\mathcal{L}|$ matrix  whose generic entry  $w_{q,r}$ indicates the weight to be assigned to edges from layer $l_q$ to layer $l_r$, a weighted flattening of $G$ is a weighted graph $(V_f,E_f,\omega)$  where $(V_f,E_f)$  is a basic flattening of $G$ and $\omega(a_i,a_j)=\sum_{\{ ((a_i,l_q),(a_j,l_r)) \in E\}}w_{q,r}$.

An alternative to complete flattening is proposed by De Domenico et al.~\cite{DeDomenico2015}, by means of a layer aggregation technique which aims at finding a compromise between the original multilayer network and a complete flattening.
The idea is to aggregate most \textit{similar} layers, based on the assumption that some layers may contain redundant information, i.e., they may show a similar topology. 
The proposed technique is based on an agglomerative hierarchical clustering schema, where the leaves of the dendrogram are the original layers, and the root is the flattened graph (and intermediate levels represent consecutive layer aggregations). Jensen-Shannon divergence is used to measure distance between layers, while the quality of each aggregation is evaluated based on information loss, measured using Von Neumann entropy. The best solution is chosen as the one corresponding to the level of the dendrogram containing the aggregation minimizing the information loss, thus providing an aggregated multilayer network representing the best trade-off between the extent of the simplification (i.e., number of aggregated layers) and information loss.

Since the main objective of flattening is that to allow the applicability of single-layer techniques on multilayer networks, in most cases task-based flattening approaches have been proposed in relation to specific network analysis tasks, e.g., community detection~\cite{BerlingerioCG11}.

\begin{figure}[t]
\centering
\includegraphics[width=0.55\textwidth]{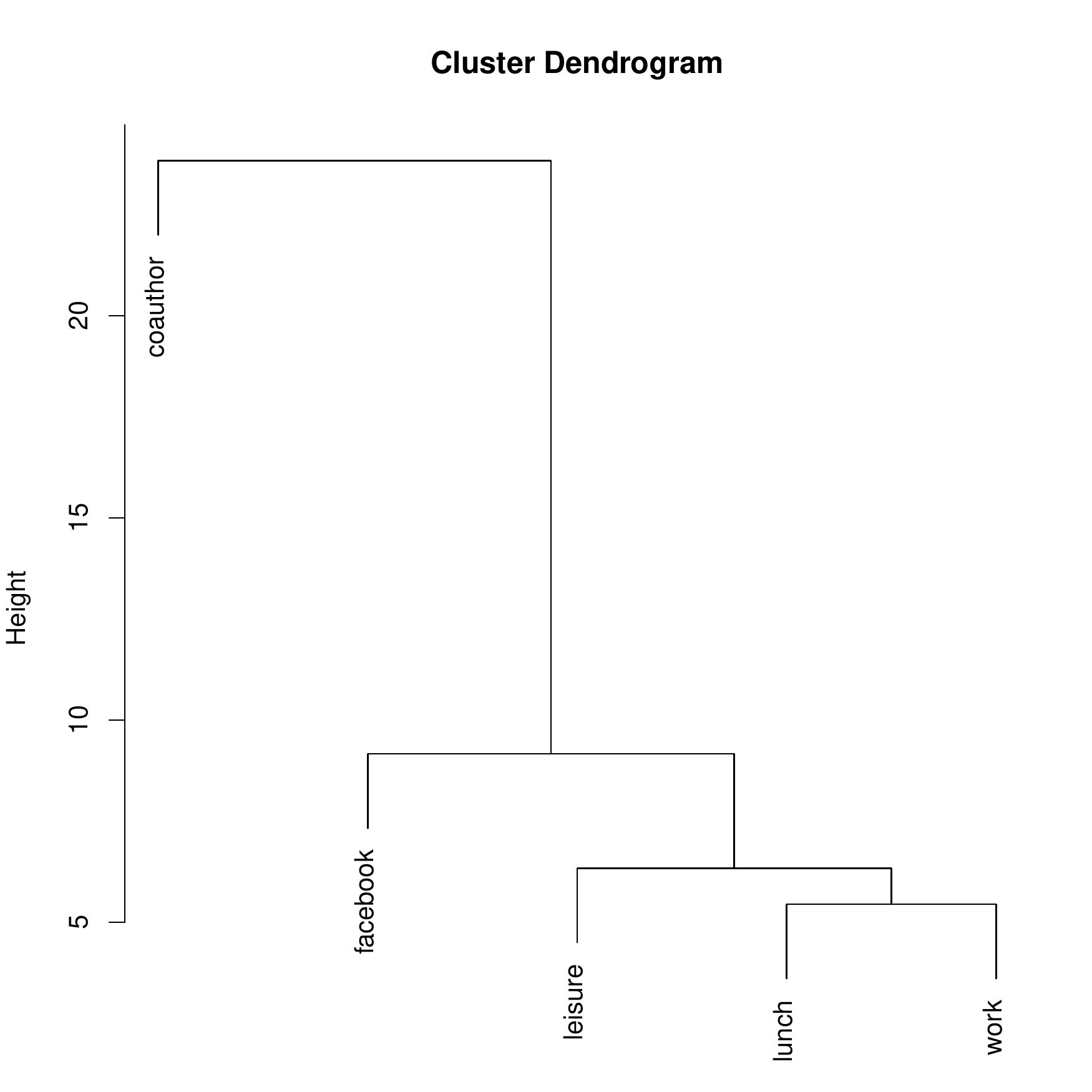}
\caption{A dendrogram showing how similar different layers are in the full AUCS network. This type of information can be used to decide which layers to merge, and this dendrogram has been computed using the inverse Jaccard coefficient to measure the portion of edges present in both layers, for each pair of layers.}
\label{fig:example_flattening} 
\end{figure}

The choice of which layers to merge (in this case, flatten) is often based on domain knowledge.  Figure~\ref{fig:example_flattening} shows how some implicit information about the similarity between different layers is however already present in the data: several layer comparison measures are presented and compared in~\cite{Brodka2017}.

\subsubsection{Multilayer community detection}
\label{sec:gpart}
Discovering an organization of the network into   densely-connected subgraphs, i.e., clusters or communities, naturally achieves a common way of simplifying a graph in terms of mesoscopic structural features due to group properties. 
Clearly, as a consequence of the learned community structure, the resulting communities could also be  used to select subgraph(s) from  the original network in order to support focused tasks.

In recent years, the problem of identifying communities has gained attention and several community detection methods specifically conceived for multilayer networks have been developed in the last few years~\cite{KimL15,Loe15,Kivela+14}.   
Community detection methods for multilayer networks can be broadly grouped into three categories, namely: flattening-based~\cite{Dickison2016,BerlingerioCG11},  
 layer-by-layer~\cite{BerlingerioPC13,pmm,TangWL12}, and direct (or multilayer)~\cite{PhysRevX.5.011027,KunchevaM15,AfsarmaneshM16}. The first group of methods are commonly used to flatten a multilayer network to enable the use of abundantly available single-layer community detection methods. 
The second category consists in processing each layer of the multilayer/multiplex network separately, and then aggregating the results. The last class of methods operates directly on the multilayer network model.

 \textit{Layer-by-layer} can be divided into three branches:  \emph{pattern mining}~\cite{BerlingerioPC13}, \emph{matrix composition}, and \emph{consensus matrix}. Pattern mining detects communities in each layer separately using a simple-graph community detection, then makes use of pattern mining algorithms to aggregate the resulting communities. Matrix-composition-based methods~\cite{pmm,TangWL12} extract  structural features from each dimension of the network via modularity analysis, and then integrate them all to find out a robust community structure among actors. The latter group of methods, consensus-matrix-based methods, combine multiple solutions over the various layers  to infer a single community structure that is representative of the set of layer-specific community structures. 
 
 The \textit{multilayer} group of methods includes \emph{clique}-based methods~\cite{AfsarmaneshM16}, which exploit the concept of multilayer cliques to identify multiplex communities, \emph{random walk}-based methods~\cite{PhysRevX.5.011027,KunchevaM15},  which introduce a multilayer random walker that can traverse inter-layer edges, \emph{modularity}-based methods~\cite{Mucha10,CarchioloLMM10}, which define a multilayer modularity function and optimize it to produce the community structure solution, \emph{label propagation} methods, which utilize a multilayer affinity measure among actors given their connections   on different layers and then introduce  a labeling method for the actors controlled by these affinity scores, and \emph{within-group connectivity} local methods, which define a multilayer within-group connectivity function for the multiplex community and tries to maximize that function. 
 
 In addition to the above categorization, we can identify an extra group of methods under the node-centric (or local) name. These query-dependent methods are designed to discover local communities starting from a group of seed nodes, and result particularly useful when global information on the whole network are missing.

\begin{figure}[t!]
\centering
\includegraphics[width=0.45\textwidth]{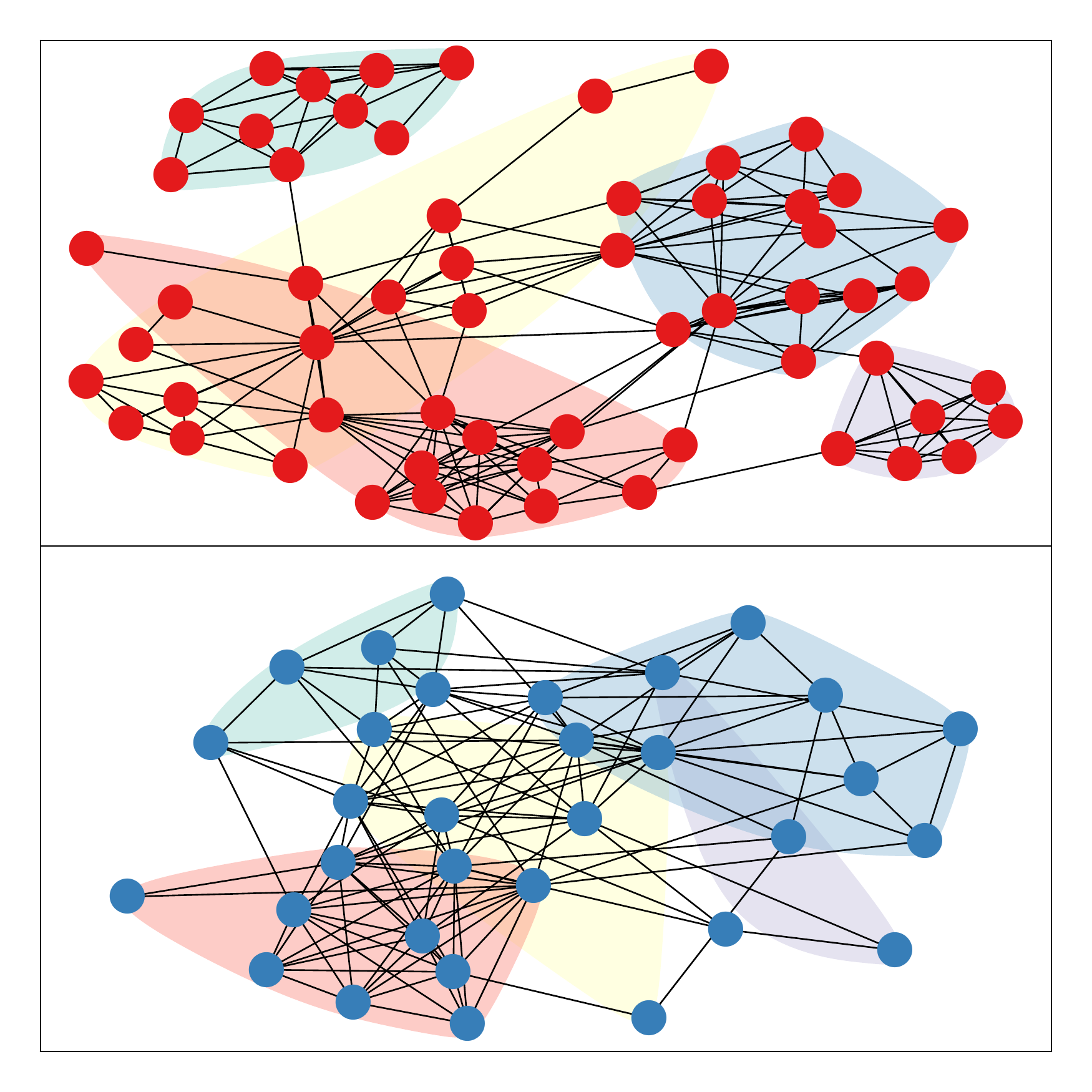}
\caption{Communities in the AUCS network. Each community can be replaced by a single node in each layer, with weighted edges between the new nodes to indicate how many edges (or which percentage of edges) were originally present between nodes from the two corresponding communities}
\label{fig:example_community} 
\end{figure}

Figure~\ref{fig:example_community} shows the communities identified by the generalized Louvain method in the AUCS network. In this case, each node belongs to exactly one community, that is, the communities define a partitioning of the node set. One can thus easily created a new network with only five nodes on each layer, to study the relationships between communities instead of the relationships between individual actors.

\subsubsection{Coarsening} \label{sec:coarsening} 
The concept of \textit{graph coarsening} (sometimes referred to as \textit{contraction}~\cite{Geisberger2012,Basak2016}) refers to a family of simplification approaches aimed at building a hierarchy of successively \textit{coarsened} (i.e., aggregated) graphs with decreasing size \cite{Buluc2016}. 
The general idea is to enable the application of computationally expensive algorithms by running them on the coarsened graph, then obtaining a solution for the original graph through an \textit{uncoarsening} (i.e., expansion) phase.

For simple graphs, a basic coarsening approach consists in contracting two nodes (i.e., a single edge) at each level of the hierarchy. Though being extremely simple, this method tends to result in a hierarchy with as many levels as the nodes in the original graph, and for this reason is referred to as \textit{n}-level approach~\cite{Osipov2010}.
Another well know coarsening strategy is based on the contraction of matchings (i.e., pair of nodes connected by edges non-incident to each other), which leads to a logarithmic number of levels. As a general rule, a good matching should contain  high weighted edges, and relatively uniform node weights. This trade-off can be handled using an \textit{edge rating} function~\cite{Holtgrewe2010,Sanders2011,Safro2015}, and specific solutions have also been proposed for scale-free complex networks~\cite{Glantz2016}. 

The typical application of coarsening is in graph partitioning, where the combination of coarsening and partitioning techniques is referred to as \textit{Multilevel Partitioning}~\cite{Bui1993,Hendrickson1995,Karypis1998,LaSalleK15}, where the idea is to exploit the multilevel hierarchy of contracted graphs produced during the coarsening phase in order to efficiently obtain an initial partitioning, which is then refined during the uncoarsening phase in order to obtain a solution for the original graph.    
Coarsening-based multilevel partitioning methods are often used to address \text{graph drawing} problems. In fact, classic force-directed graph drawing algorithms~\cite{Fruchterman1991},  when dealing with large graphs, can easily get stuck in a local minimum.
Using a multilevel approach allows to find global optimal layouts for (relatively small) graphs in the lower levels of the hierarchy in reasonable time, which are then used to build a solution for the original graph through an iterative refinement process.
Several graph drawing algorithms based on this multilevel approach have been proposed, which are able to produce high-quality drawings for large graphs, i.e., scalable up to millions of nodes~\cite{Walshaw2003,Hachul2005,hu2005efficient,Martin2011}.

To best of our knowledge, no specific coarsening techniques have been proposed for multilayer network models. An intuitive reason can be that contracting adjacent nodes and edges can be naturally thought as a \textit{layer-wise} action, so that taking into account different layers at the same time may be redundant or misleading. 
In other cases (i.e., where the source of complexity lies on the number of layers) layer aggregation approaches may be preferred in order to solve similar problems (cf. Section~\ref{sec:flat}).
Nevertheless, we believe that native multilayer coarsening approaches can be envisaged based on the contraction of inter-layer edges and/or exploiting multilayer information about a node's neighborhood (e.g., leveraging on relevance-based measures~\cite{BerlingerioCGMP13}). Such techniques may also help to advance research on multilayer network visualization, which is still a challenging problem~\cite{cnu038}.

\subsubsection{Positional equivalence}
\label{sec:poseq}

Another common approach to analyze social networks is to study their \emph{structural roles} and \emph{structural positions}. That is, finding actors, or groups of actors, whose local connectivity in the graph define their role in the social network. This idea, known as positional equivalence, was originally developed by the sociologists Lorrain and White in the early 70s~\cite{lorrian1971structural:methods}, under the assumption that actors related to exactly the same other actors in the network must share an equivalent social behaviour.

The main difference between both structural entities, positions and roles, has been largely discussed in the literature. The definition provided by Wasserman and Faust~\cite{Wasserman1994}~\footnote{The terms \emph{role-equivalent} and \emph{role-similar} are also commonly used to refer to the  original notions  of \emph{role} and \emph{position}} emphasizes that \textit{structural roles} refer to actors with similar patterns of connectivity, independently of the specific actors to whom they are connected (e.g., hubs in a network); while \textit{structural positions} are based on the concept of interchangeability: two actors in the same structural position can be swapped without changing their relationships with other actors in the network.  

The most straightforward and strict type of equivalence is  \emph{structural equivalence}~\cite{Faust1988}, which places nodes in the same position if, and only if, they have the same connections (i.e., the set of adjacent nodes is identical). This type of equivalence is difficult to find in practice in large and/or complex networks and can be replaced by \emph{regular}~\cite{LUCZKOVICH2003}, \emph{automorphic}~\cite{Everett1991} and \emph{stochastic}~\cite{Holland1981} definitions of equivalence, which are able to find more flexible types of positions. In short, the stochastic equivalence definitions assign nodes of the graph to the same role if they have the same probability distribution of edges with other nodes.

The most common and extended methods to detect roles in social networks are \textit{block-modeling}~\cite{ryan2015roles:sna} methods, which allow   the representation of a network using an image matrix where the nodes, grouped into blocks, represent structurally equivalent roles or stochastically equivalent positions, and the edges represent interactions between them. In the past years, new block-modeling methods has been developed to find roles in complex graphs such as two-mode~\cite{Doreian2004} and multilevel~\cite{ZIBERNA201446} networks. One of the main applications of stochastic blockmodels and similar generative methods is to detect network communities, as the image matrix represents --- under certain conditions --- a partition of the network that maximizes the posterior modularity coefficient of the partition~\cite{peixoto2017bayesian,peixoto2018nonparametric}. Many advancements in the field have been limited to finding community structures restricted to particular layers~\cite{Stanley2016} or pillars~\cite{paul2016consistent}. In a recent paper, Pamfil et al.~\cite{pamfil2018relating} proposed a block-modeling method to detect different types of structural roles and positions for various types of multilayer structure, including temporal, multiplex, and multilevel (i.e., hierarchical) networks.

While block-models are the most popular methods to find positional equivalences, there are also other methods to compute structurally equivalent or similar roles based on similarity measures (e.g., Euclidean distances, correlations) computed on the structure of the nodes of the graph and/or their attributes~\cite{Henderson2012}. However, to the best of our knowledge, in multilayer networks these methods have been only used in combination with some of the aforementioned block-modeling methods~\cite{peixoto2018nonparametric}.

\subsubsection{Graph compression and summarization} \label{sec:compr}

Graph compression and summarization refer to a set of simplification techniques created in order to improve the volume and storage of the network, usually with the goal of speeding-up graph algorithms, queries~\cite{Fan2012} or the visualization of large graphs~\cite{Maneth2015ASO,LiuSDK18}. While both concepts have been used interchangeably, they differ on their main objective. In general, summarization  or \emph{semantic graph compression} techniques focus on compressing structural features of the graph that have a semantic in a specific application domain, usually for visualization purposes. The goal of these methods, therefore, is not very different from the goal of the sampling and coarsening methods we have introduced above, and hence we will not review them here.

\emph{Algorithmic graph compression} aims instead  to loosely reduce the graph size for an efficient execution of graph mining tasks~\cite{Fan2012,Toivonen2011,Lim2014,Khan2016,Nourbakhsh2014}. The perfect graph recovery is sometimes unfeasible, therefore most of the recent advancements in graph compression have focused on developing faster algorithms that guarantee an acceptable recovery rate~\cite{MANETH201819}.

\begin{figure}[t!]
\centering
\includegraphics[width=0.9\textwidth]{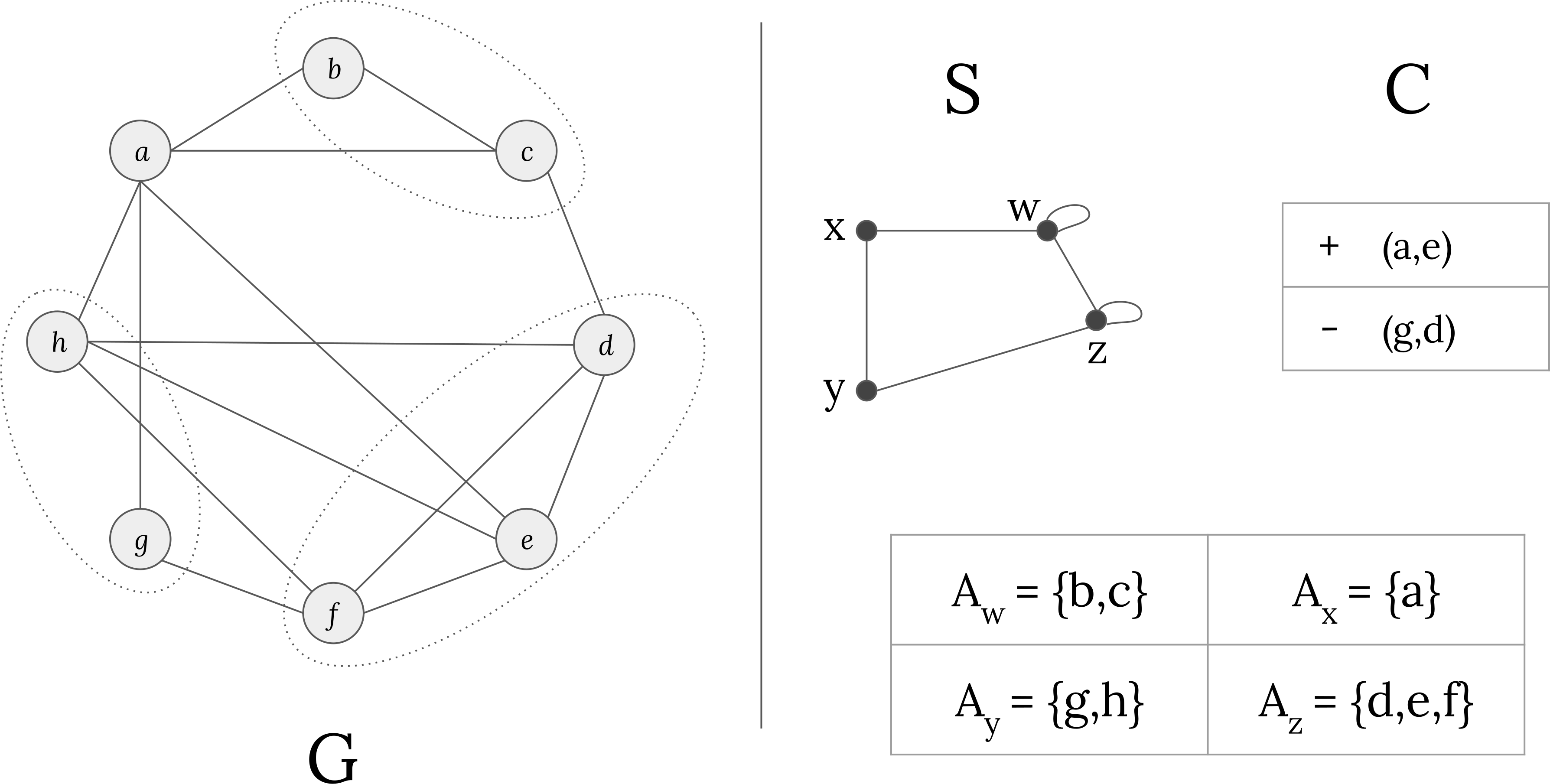}
\caption{The two-part MDL  representation proposed in~\cite{Navlakha2008}: original graph ($G$),  graph summary ($S$), corrections ($C$), and the super-node mappings.}
\label{fig:compression}
\end{figure}

Most of the algorithms for graph compression are based on the ideas from Navlakha et al.~\cite{Navlakha2008}, who proposed using a two-part Minimum Description Length (MDL)~\cite{Rissanen1978} codification to represent the simplified network. The idea is to represent the original graph $G = (V_G , E_G )$ using two components: an aggregated or coarsened graph $S=(V_S, E_S)$ and a set of corrections $C$. The summary $S$ is a simplified network with considerably fewer nodes and edges   that can easily fit in memory, while the corrections are two sets representing the edges that should be added or removed to $S$ in order to recover the original graph. The size of $C$  depends on the specific structure of the original graph $G$ and on the level of compression applied. Figure~\ref{fig:compression} shows an example where the original 8-node graph is  compressed into a summary of 4 super nodes.

Several proposals appeared recently to deal with attributed nodes and edges, allowing to compress weighted~\cite{Khan2016,Toivonen2011} and temporal networks~\cite{Shah2015}. However, a difficult research question left unanswered in the literature is how to adapt graph-compression techniques to complex graph structures, such as   multilayer networks~\cite{LiuSDK18}. While most of the coarsening and sampling methods for multilayer networks described in Section~\ref{sec:coarsening} and Section~\ref{sec:sampling} could potentially be used to generate a summary of the graph, researchers have not found yet a correction algorithm or structure able to simultaneously recover the original network while reducing the representation size of the graph in memory.

Methods based on node grouping, for example, can efficiently reduce the size of the intra-layer graphs, but this will not reduce the amount of storage as we still need to keep track of the inter-layer edges within the original nodes across several layers. On the other hand, while structurally equivalent positions   can be seen as graph summaries with no errors, they are extremely rare in real networks, which makes them unsuitable for graph compression.

\subsection{Transformation}

All simplification methods result in the removal of some objects, e.g., edges or nodes, from the network. In the previous two sections we have discussed approaches that just remove some objects, without replacing them with anything else (selection) and other approaches replacing the removed objects with less objects of the same type, e.g., replacing groups of nodes with a single one (aggregation). The final main class of simplification methods shares with the previous two the fact that some objects are removed from the network, but in this case the removed objects are replaced by (that is, transformed into) objects of a different type.

\subsubsection{Projection-based}
\label{sec:proj}

The term \textit{network projection} generally refers to simplification techniques dealing with two (or more) node types in a network. The traditional approach consists in reducing the number of node types by replacing edges between two node types with edges between the same types of nodes, and removing one of the node types.

For instance, given a two-mode network with node types $A$ and $B$, the typical projection approach consists in removing type-$B$ nodes and adding an edge between any pair of type-$A$ nodes originally connected to the same type-$B$ node~\cite{Seierstad2011}.
As a practical example, a two-mode co-authoring network containing \textit{author} and \textit{publication} nodes can be simplified by inserting an edge between two author nodes when they are both connected to at least a common publication node, thus discarding the \textit{publication} type.

Even if we do not explicitly take into account  networks including heterogeneous node types in this work, in a multilayer context it is easy to think about the same process for nodes belonging to different layers, i.e., nodes belonging to two different layers $l_a$ and $l_b$ can be thought of as nodes of two different types $A$ and $B$ (i.e., the \textit{author} and \textit{publication} node types in the previous example can be easily modeled as two different layers in a multilayer network).

Nevertheless, the standard projection approach described in the previous example shows several shortcomings. First of all, all the information regarding multiple connections between the same pair of nodes would be lost, e.g., number of common publications between two authors. Moreover, it tends to generate large cliques, especially in presence of extremely popular \textit{hub} nodes of the type being discarded. Padrón et al.~\cite{Padron2011} introduced a weighted projection technique which tries to overcome these limits by weighting the edges with the number of common connections from which they are originated. More advanced weighting schemes are proposed in~\cite{Newman2001,Opsahl2013}, e.g., based on the concept of network relevance and with weights redistributed according to the exclusivity of the collaborations. 

\begin{figure}[t!]
\centering
\subfigure[]{\includegraphics[width=0.45\textwidth]{./imgs/original-2mode.pdf}} 
\subfigure[]{\includegraphics[width=0.45\textwidth]{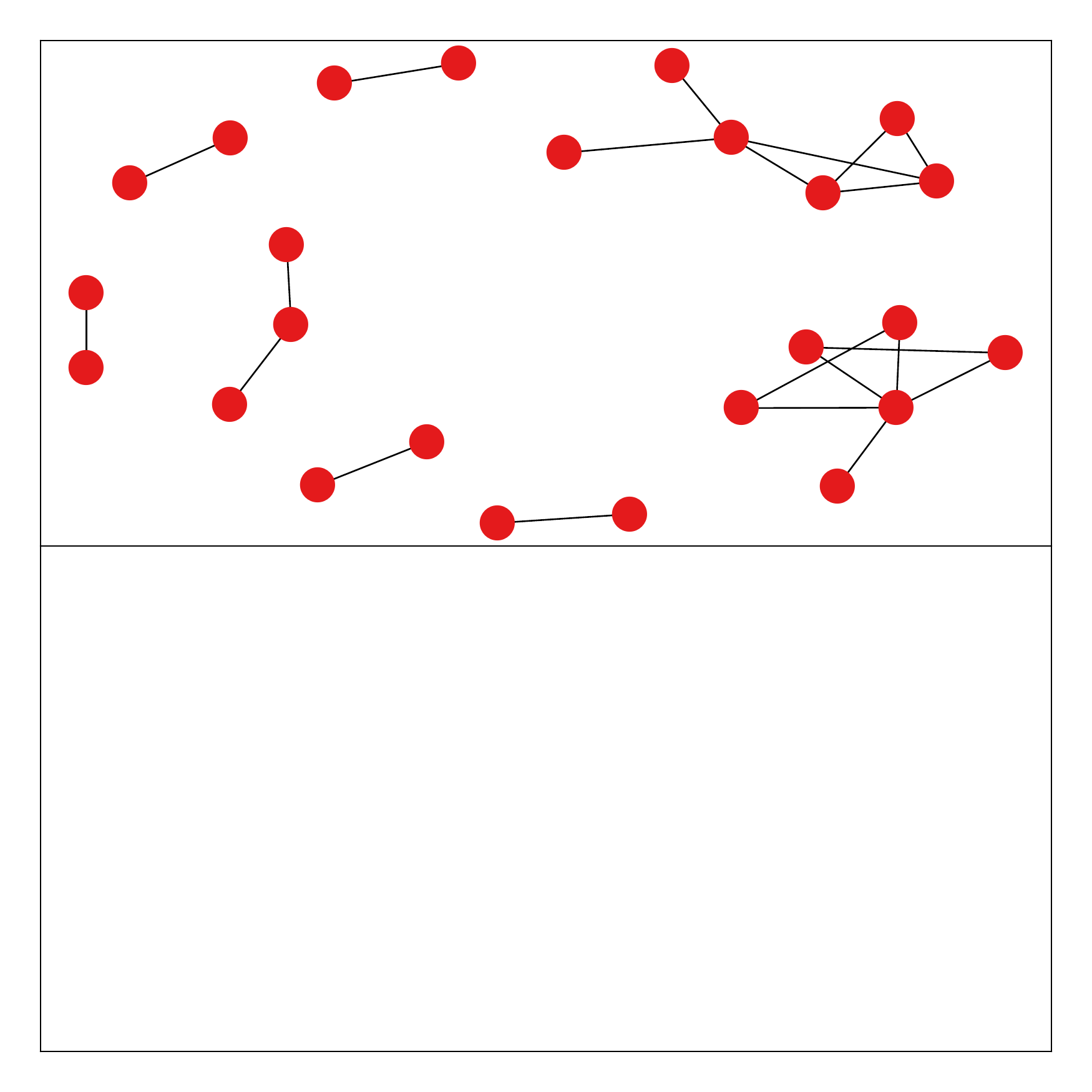}} 
\caption{(a) Author and paper layers in the AUCS network. (b) A projection of the papers on the author layer.}
\label{fig:example_projection} 
\end{figure}

Figure~\ref{fig:example_projection} shows how a projection can be used to create a simple network (or a simple layer) starting from two layers with different node types and inter-layer edges.

\subsubsection{Graph-embedding-based}
\label{subsec:embedding}
Graph embedding techniques   aim to transform a graph in a low-dimensional representation, enabling the use of a rich set of tools mostly based on state-of-the-art machine-learning methods~\cite{cai2018comprehensive}.  
 A clear advantage in obtaining a low-dimensional representation is to reduce the memory-footprint requirements while retaining relevant information for the task at hand~\cite{bengio2013representation}. Example tasks include node classification, node clustering, node recommendation, retrieval and ranking, link (i.e., edge) prediction,  and graph classification~\cite{GoyalF18}. 
  Moreover, since the embedding corresponds to a vectorial representation, vector operations on the learned model might in principle be     computationally more convenient than graph operations.  
Another task which benefits from graph embedding techniques is network visualization. Given the rising complexity and volume of current networks, graph embedding techniques represent an essential tool to reduce the size of the network, capture meaningful patterns and thus build visual representations which can easily convey properties and structural information of complex networks.    
Traditional visualization algorithms seek to find a compromise between clarity of structural characteristics and aesthetic requirements such as fixed edge lengths or minimal edge crossing (e.g., \cite{kk89, pt98, dbd13}) while resorting to \textit{multidimensional scaling} to find a low-dimensional representation of high dimensional data, with the goal of preserving pairwise dissimilarities in terms of Euclidean distance~\cite{bg05,bg13}.

The general goal of a graph embedding process is to learn a function $f$ that maps one or many features   of the network (i.e., nodes, edges, or the whole graph) to a new $d$-dimensional space $\mathbb{R}^d$. 
 For instance, in the case of node embedding for a multilayer network, the function to learn can be of the form   $f:\mathcal{A} \rightarrow \mathbb{R}^d$ or $f:V \rightarrow \mathbb{R}^d$.  
 In effect, graph embedding approaches can be classified based on the constituent(s) of a graph they are designed for. Several works have been recently developed for  
  nodes~\cite{deepwalk,node2vec}, edges~\cite{ZhaoZWHC16}, subgraphs (e.g., communities~\cite{WangCWP0Y17,CavallariZCCC17}), and whole-graph embedding \cite{0001KM17,NiepertAK16,MousaviSMB17}.

\textit{Matrix factorization} based algorithms~\cite{HofmannB94,HanS16,YinGL16} are the pioneering methods in graph embedding, since they apply a decomposition technique   to a matrix representation    of an input graph.  There are mainly two types of matrix-factorization-based graph embedding: factorization of graph Laplacian eigenmaps, and direct factorization of  the node proximity matrix~\cite{cai2018comprehensive}.
 One of the earliest methods  designed for multilayer networks, specifically for community detection purposes, is  Principled Modularity Maximization (PMM)~\cite{pmm}.   PMM  infers a latent community structure for the nodes in a multilayer network by performing a two-step methodology. Firstly, PMM extracts low dimensional vectors from each layer through modularity maximization and aggregates the extracted information through cross-dimension integration. Finally, a simple k-means is carried out on the learned representation to find out the communities of the network.
 
 From a different perspective, 
in the LINE method~\cite{line},  two functions are defined  for both first- and second-order proximities, where first-order proximity refers to edge weights and   second-order proximity refers to neighborhood similarity.    LINE defines two joint probability distributions for each pair of nodes  and minimizes the Kullback–Leibler (KL) divergence of these two distributions.

More recently, there has been  momentum for the development of \textit{deep-learning-oriented} techniques. 
DeepWalk~\cite{deepwalk} and  Node2Vec~\cite{node2vec} are two exemplary random-walk-based methods. According to the \textit{Skip-gram}~\cite{mikolov2013} model, these methods treat nodes as words and paths as sentences, then apply deep learning to the sampled random-walk paths.  
As the skip-gram model aims to maximize the co-occurrence probability among the words that appear within a window in the same sentence,  the resulting graph embedding by DeepWalk and Node2Vec preserves first- and second-order proximity of nodes.

Other deep-learning-oriented methods  include the use of  
autoencoders and their variants (i.e., denoising, variational, etc.), which  aim to maximize the reconstruction accuracy of the input graph, by applying a \textit{decoder} block to the latent representation learned by an  
\textit{encoder}~\cite{sae,dngr,sdne}.  
Another deep-learning common approach is to directly apply a convolutional neural network (CNN) to Euclidean data generated from a graph~\cite{NiepertAK16}, or to adapt  CNN  to graphs~\cite{gcn, scarselli2009graph}. In addition to this group of methods, GraphSAGE~\cite{graphsage}  aims to learn a function that generates embeddings by sampling and aggregating features from a node’s local neighborhood. GraphSAGE, which can be seen as an extension of the GCN~\cite{gcn} framework to the inductive setting,  can deal with  evolving graphs  and it can easily generate embedding vectors for previously unseen nodes.

\paragraph{Graph embedding for multilayer networks} 
 At the time of writing of this paper, there is a relatively small corpus of   
 methods of graph embedding specifically conceived for  multilayer networks, namely: PMNE~\cite{PMNE}, MNE~\cite{MNE}, and MELL~\cite{MELL}.

\begin{figure}
\centering
   \includegraphics[width=0.9\linewidth]{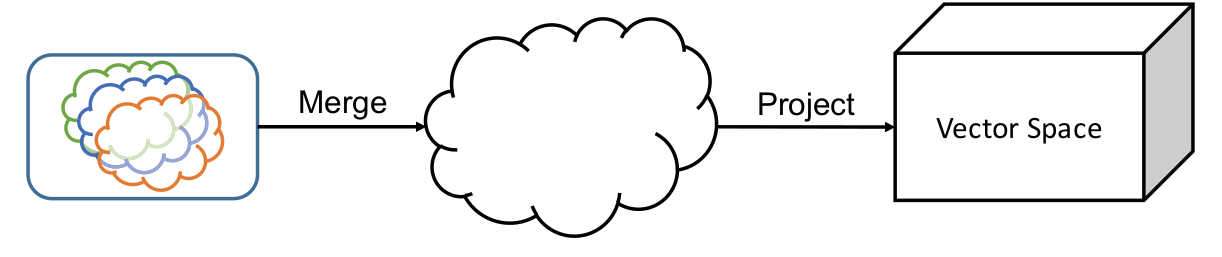} \\
   (a) \\ 
   \includegraphics[width=0.9\linewidth]{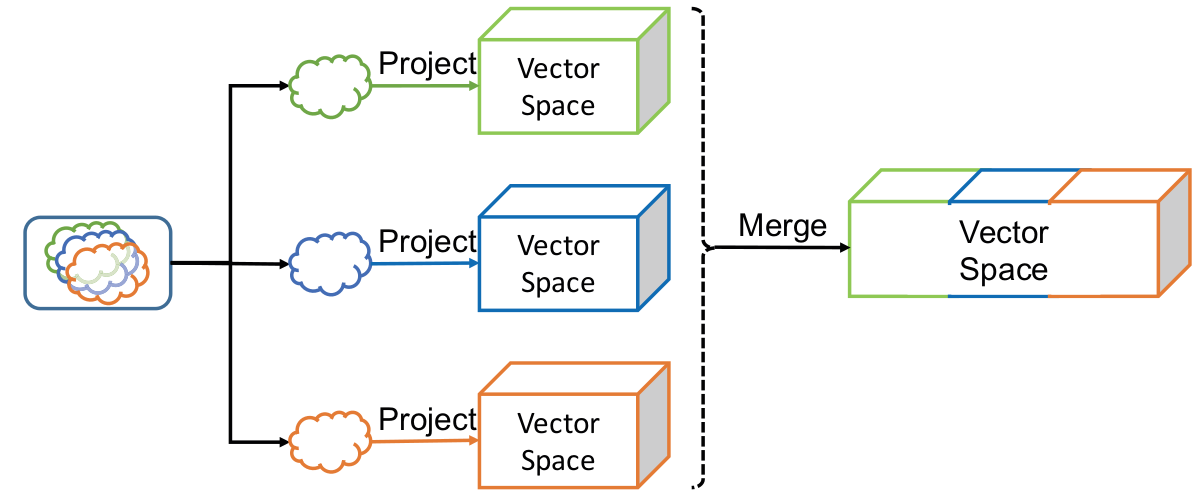} \\
   (b) \\ 
   \includegraphics[width=0.9\linewidth]{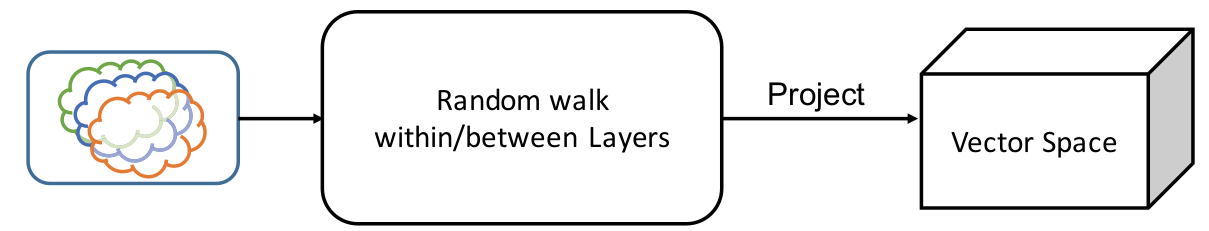} \\
   (c)  
\caption{Architecture of PMNE: (a) Network Aggregation, (b) Results Aggregation, (c) Layer Co-analysis~\cite{PMNE}.}
\label{pmne}
\end{figure}

The first two methods adapt and extend the Word2Vec~\cite{w2v1,w2v2} model  to multilayer and multiplex networks, respectively. The input  sentences (i.e., paths) are generated by a second-order random walk process, which is constrained to explore one layer at the time, with the exception of PMNE Layer co-analysis method in which the random walker gains the ability to jump from a layer to another.  
As depicted in Fig.\ref{pmne}, PMNE includes three different approaches: two naive baselines, and one natively multilayer. In the Network Aggregation approach,   the multilayer network is merged into a single weighted network (where multiple edges between two nodes are not allowed) and the embeddings for actors are computed on the aggregated network, i.e., $f:\mathcal{A} \rightarrow \mathbb{R}^d$.  
 In the Results Aggregation approach, the embeddings are computed separately on each layer and then successively concatenated together, i.e., $f_l:V \rightarrow \mathbb{R}^{d_l}$ and $f=f_1 \lvert \rvert f_2 \lvert \rvert \cdots \lvert \rvert f_{\lvert L \rvert}$ with $f:V \rightarrow \mathbb{R}^{d'\lvert L \rvert}$. 
 Unlike the previous two approaches, in the Layer Co-analysis approach the second-order random walker acquires the ability to jump across layers, allowing the generation of walks that are not limited to a single layer of the network, i.e., $f:V \rightarrow R^d$. 
 In all of the three approaches,  the  Node2Vec model is chosen to be  applied on the generated paths. 
 
MNE generates a layer-wise embedding $\textbf{v}_n^i$, for every node $n$ and layer $i$, which consists of  an embedding $\textbf{b}_n$ shared across all the layers and that describes the node $n$ globally, and a local (i.e., layer-wise) embedding $\textbf{u}_n^i$: 

\begin{equation*}
\textbf{v}_n^i = \textbf{b}_n + w^i \cdot {\textbf{X}^i}^T \textbf{u}_n^i.
\label{mneq}
\end{equation*}
In the above equation,  $w^i$ denotes the importance of layer $i$, and ${\textbf{X}^i}^T$ is a transformation matrix that aligns the global and local embedding vectors.

MELL is based on a regression framework, and unlike the previously described methods, it also takes into account the directionality of the edges (i.e., $(u,v) \neq (v,u)$) by using two vectors for each node $\textbf{v}_H$ and $\textbf{v}_T$. Node embeddings belonging to the same node are enforced to be close to each others through a regularization term. In addition to node embedding, MELL also learns a set of \textit{layer vectors} representing layers' connectivity, in order to differentiate edge probabilities in each layer. 
The probability between two nodes $v_i^l$ and $v_j^l$, both belonging to the layer $l$, is equal to:

\begin{equation*}
q(\textbf{v}_i^l, \textbf{v}_j^l) = \frac{1}{1+\exp \left(-(\textbf{v}_i^l + \textbf{r}_l)^T \cdot \textbf{v}_j^l \right)}
\end{equation*}

\noindent
where $\textbf{v}_i^l$ (resp. $\textbf{v}_j^l$) denotes the embedding vector for $v_i^l$ (resp. $v_j^l$), and $\textbf{r}_l$ is the vector embedding for the $l$-th layer.  

Both MELL and MNE are well-suited to link prediction and node classification, where high-order proximity information, extracted through random walk, turns out to be particularly expressive. MELL  could be preferred when the information carried by edge directionality is essential for the task at hand. Conversely, MNE might in principle be more suited to exploit both   global  and  local node-embeddings.

\section{Discussion}
\label{sec:discussion}

Building upon our analysis in the previous sections,  here we provide a few remarks that are   concerned with the following  two questions:   
{\bf (RQ1)} What are the main characteristics that would make a given     approach appropriate or not   for a given   simplification task?
and, {\bf (RQ2)} How is research on network simplification going to evolve, given the existing corpus of simplification methods?  

Concerning {\bf RQ1}, we will focus on    practical usability criteria to determine whether a simplification approach is appropriate or not for a problem at hand, namely  the parts of the network that are affected, whether the result is deterministic, and whether  the simplification process is reversible (cf. Table~2).   
 It should also be noted that the variety of network simplification tasks and, consequently, the diversity in their respective analysis purposes makes it difficult to devise a unified framework for the evaluation of network simplification methods. Ultimately, this  could be useful to support the choice of a specific algorithmic solution among different alternatives.  
For simple graphs, we can   
recognize a number of evaluation aspects that might be meaningful and relevant for any of the network simplification approaches; these aspects certainly include the opportunity of measuring, from one or more perspectives (i.e., micro-, meso- and macroscopic),  the extent to which the simplified network preserves selected structural properties in the original network and whether other structural properties are changed.
 Estimating the sampling bias is clearly a crucial aspect for methods in the selection-oriented simplification category, while aspects relating to information loss may also be of interest to evaluate the quality of transformation- and aggregation-oriented methods.  
Compression ratio and related concepts, with their impact on the spatial complexity, might represent useful indicators for memory-footprint requirements of the simplification method.

Practical usability is also related to the availability of software implementations. According to our findings, several simplification methods for multilayer networks are still to be invented.  
 Therefore, software support for multilayer networks in general, and simplification approaches in particular, is still more limited than what is available for non-multilayer networks. However, some of the methods reviewed in this article come with freely available software implementations, and some are also available as part of multilayer network analysis libraries, either as single function calls or requiring some limited coding. In Section~\ref{sec:tools}, we review these implementations.

  \subsection{Practical usability aspects}
  
    \subsubsection{Key-enabling and Affected elements}

Even though multilayer networks are defined over a richer set of elements, it can be noted from Table 2  that most selection- and aggregation-oriented methods rely on (and affect) basic network elements, i.e., nodes (set $V$) and edges (set $E$). Moreover, all methods aim at reducing the number of nodes and edges in the network. 
This is not surprising, since most techniques represent adaptations to multilayer models of methods which were designed for single-layer networks.  
As a matter of fact, the main exceptions in this scenario are represented by the families of methods  that are specifically tailored to multilayer networks, i.e., flattening and layer aggregation.
A different relation between key-enabling and affected elements can be observed for transformation-oriented methods. Concerning projection, this is the only case where    increase in an affected element ($E$) can be observed.  A slightly different situation occurs for embedding-based methods, where a  new edge set is derived from a certain embedding space (over nodes or actors), which is forced to be  
 smaller in size with respect to the original one. 
Summarizing, while selection- and aggregation-oriented methods always aim at a simplification which results in a reduction in size of the original network, transformation-oriented methods rely on a different concept of simplification (i.e., suppressing a node type for projection, and obtain a low-dimensional representation of the network in the case of network-embedding).

 \subsubsection{Determinism} 
Looking at the fifth column of Table~\ref{tab:cat}, it can be noted that selection-oriented 
simplification methods can be non-deterministic, especially in sampling approaches. These   inherently rely on non-deterministic   techniques that aim to produce a subnetwork (sample) exhibiting similar properties as compared to the original network, whereby nodes or edges are randomly picked to start  the building of the sample. 
Also, in model-based filtering, non-determinism may arise due to the sampling of ensembles of networks according to constraints on degree, strength and other properties based on probabilistic generative models.   
 
 Aggregation-oriented simplification  mostly  includes   deterministic methods, provided that no    randomization or low-rank approximation techniques are used, as it might be the case for coarsening and  multilevel-partitioning (where the hierarchical approximation obtained via coarsening is sequentially refined over  all levels of the hierarchy  until it accomplishes the task for the original input with some approximation for it), and for some community-detection methods such as, e.g., label propagation, whose uniqueness of solutions is not ensured.

 Transformation-oriented simplification techniques are deterministic when dealing with network projection tasks, whereas embedding-based transformation should be regarded as   non-deterministic when low-rank approximation techniques are used to compute the embeddings.

\vspace{3mm}
  \subsubsection{Reversibility}

Concerning reversibility, a general remark is that simplification   methods are usually  designed   without any particular requirement in terms of ability to reconstruct the original network from a simplified one, and hence simplification is typically interpreted as irreversible. For instance, this is the case of multilayer community detection, whereby a task of aggregation-oriented simplification would in principle be carried out over the output community structure without storing the full topology of the original network. 

Methods whose outputs could be reversed include   multilevel partitioning, positional equivalence, graph compression, and coarsening, as long as the existing methods are natively equipped with procedures that enable reversibility or recovery  of the simplification process; however, none of such   reversible approaches has a multilayer counterpart. 

This would  raise the opportunity for developing  reversible methods, in particular for compression and  coarsening, that are specifically conceived for multilayer networks.  
 In particular, it would be interesting to investigate the minimal requirements in terms of network information   that needs to be indexed and stored, in order to reconstruct the original graph, or infer it with a bounded approximation guarantee, in relation with existing simplification techniques.

 \begin{landscape}
\begin{table} 
    \caption{Available implementations of native methods for  multilayer network simplification.} 
    \label{tab:impl}
    \centering
    \scalebox{0.95}{
        \begin{tabular}{|l|l||cl|}
        \hline \hline 
        \textbf{category} & \textbf{method} & \textbf{Refs.} & \textbf{Public repository} \\
\hline

\multirow{5}{*}{Selection} &  \multirow{2}{*}{centrality-based filtering} & \cite{Sole14} & https://github.com/manlius/muxViz \\
    & & \cite{GalimbertiBG17} & https://github.com/egalimberti/multilayer\_core\_decomposition \\ \cline{2-4}
	&     node-layer relevance filtering & \cite{Rossi2015} & https://bitbucket.org/uuinfolab/uunet \\ \cline{2-4}
	& model-based filtering & \cite{MandaglioAT18} & http://people.dimes.unical.it/andreatagarelli/emcd/ \\ \cline{2-4}
    & sampling & -- & \\
\hline

\multirow{9}{*}{Aggregation} & \multirow{3}{*}{community-detection-based} & \cite{PhysRevX.5.011027} & https://github.com/manlius/muxViz \\
    & & \cite{AfsarmaneshM16} & https://bitbucket.org/uuinfolab/uunet \\
    & & \cite{InterdonatoTISP17} & http://people.dimes.unical.it/andreatagarelli/mllcd/ \\ \cline{2-4} %
	&	 multilevel partitioning  & \ding{55} & \\ \cline{2-4}
	&  positional equivalence & \cite{ZIBERNA201446} & https://cran.r-project.org/web/packages/blockmodeling/index.html \\ \cline{2-4} 
	&  flattening & -- & \\ \cline{2-4}
	&   layer aggregation & \cite{DeDomenico2015} & https://github.com/manlius/muxViz \\ \cline{2-4} 
 	& graph compression & \ding{55} & \\ \cline{2-4}
	& coarsening  & \ding{55} & \\ \cline{2-4} %
\hline    

\multirow{3}{*}{Transformation} 
     &	 projection-based  & -- & \\ \cline{2-4}
    & \multirow{2}{*}{embedding-based} & \cite{PMNE} & https://github.com/tangjianpku/LINE \\
    & & \cite{MNE} & https://github.com/HKUST-KnowComp/MNE \\
\hline 
\end{tabular}
}
\end{table}
\end{landscape}

 \subsubsection{Single-layer vs. Multilayer native methods}

We have noticed   that most simplification techniques correspond to adaptations or extensions to multilayer networks of methods natively designed for single-layer networks. 
However, the picture is quite variegate and contains exceptions. 
As  reported in Table~2, some of the simplification methods are only available for single-layer networks, which is the case of   graph compression and coarsening,  or no counterpart has been yet developed for multilayer networks, as for multilevel partitioning. By  contrast,  other methods are conceived for multilayer networks only, such as node-layer relevance filtering, flattening, and layer aggregation. 
Moreover,  when there is wide corpus of studies available for certain categories, i.e.,   centrality-based filtering,   our focus was  directly on the existing methods that can simplify  multilayer networks.

\subsection{Multilayer network simplification software}
\label{sec:tools}

 Table~\ref{tab:impl} records all the openly available implementations of the native multilayer methods and algorithms discussed in this review, accessed on January, 2020. Some of the aggregation methods, such as multilevel partitioning, graph compression and coarsening do not have any openly available implementation because we do not have any reference algorithm (cf. Table~\ref{tab:cat}). We have indicated them in   Table~\ref{tab:impl} with a \ding{55} symbol. For others methods, such as sampling, flattening and projection-based transformations we have one or more reference works describing a native multilayer method, but the authors have not included any reference or link to an openly available implementation. In summary, we have found that only 11 of the simplification methods in this review categorized as native multilayer contain a reference to some public available implementation or repository.

 Formally, multilayer networks are graphs where nodes are complex objects made of actors and layers. Therefore, in some cases the authors have used libraries for simple graphs to perform some operations on multilayer networks and implement some of the methods described in this article. However, none of the papers using this methodology provide a link to an open-source repository.
  
\paragraph{General libraries}
Some specific libraries are available providing native multilayer network objects and manipulation functions. In this section we focus on these libraries, and in particular \texttt{multinet},\footnote{https://CRAN.R-project.org/package=multinet, last update on git repository: January 2020, main usage with R, C++ code also available}   \texttt{muxviz},\footnote{http://muxviz.net, last update on git repository: January 2018, main usage through a visual interface}  and \texttt{pymnet}.\footnote{https://bitbucket.org/bolozna/multilayer-networks-library, last update on git repository: July 2018, main usage with python}

Selection methods are not directly available as single software functions, but some of them can be implemented using other functionalities provided by the reviewed libraries. The \texttt{multinet} library provides a selection of  centrality measures and relevance functions in addition to functions to remove nodes and edges, and their combination can easily be used to perform 
centrality-based and node-layer relevance filtering. 
The \texttt{pymnet} library provides a function to produce  induced subgraphs that can be used together with its functions to compute centrality measures  and       clustering coefficient. The \texttt{muxviz} library also provides the computation of some centrality measures, although the library is more oriented towards visual analysis than network manipulation. Model-based filtering and sampling are not currently supported, although simple sampling based on uniform probability or probabilities computed from centrality values can   easily be implemented using \texttt{multinet} and \texttt{pymnet} writing some additional R, C++ (for \texttt{multinet}) or Python code.

 Support for aggregation is limited. Both \texttt{multinet} and \texttt{muxviz} provide a selection of  community detection methods, but they do not provide a direct way to aggregate communities into single nodes. A flattening function is implemented in \texttt{multinet}, both for weighted and unweighted multilayer networks. A method to aggregate layers, also mentioned in this survey, is  available on \texttt{muxviz}, whereas  \texttt{pymnet} provides an aggregation function to reduce the number of aspects. Multilevel partitioning, positional analysis, graph compression and coarsening are not supported.

 Concerning transformation methods, a previous version of \texttt{multinet} provided a projection function, not present in the current version. None of these libraries include embedding methods, which cannot   easily be implemented just using the available functions because they require machine/deep  learning engines. 

 In summary, support for simplification methods is still very limited, although some of the approaches are available. Looking at some general libraries for multilayer network analysis, our general perception is that some of the approaches described in this article would require very little effort to be integrated in existing software, being based on basic functionality that is already available. Others, such as the approaches based on positional analysis, would require some additional effort to implement new algorithms from scratch. Methods such as network embedding may require the integration with other software providing the machine/deep learning algorithms required by these approaches. The taxonomy we introduced in this article can be used as a roadmap to guide the integration and development of simplification methods in existing libraries.

 \section{Research Directions}
  \label{sec:future}

   In relation to  {\bf RQ2} stated in the previous section, here   we focus on   the future evolution of multilayer network simplification, which is still in its infancy.  
 We shall identify  
underrepresented categories of simplification methods for multilayer networks.  
 We will highlight the most evident limitations of the existing methods, for each of the categories, and   raise   the emergence of novel classes of methods for enhancing network simplification tasks.

 \subsection{Coverage and limitations of existing approaches}
 
Selection-oriented simplification methods are relatively well-represented in the multilayer context, since for each of the approaches belonging to this category, there is at least one equivalent method conceived for multilayer networks. 
In the last two decades, we have witnessed the development of a plethora of centrality measures, mostly designed for single-layer networks. Recently, a relatively large corpus of study has addressed the extension and adaptation of single-layer centrality measures to multilayer networks  
 (cf. Section~\ref{sec:filtering}). Nevertheless, it appears that the potential of multilayer centrality methods might still be unveiled.  
  Two promising alternative approaches to the simplification process are based on  node-layer relevance and model-based filtering; however, the design of such methods for multilayer networks is still in its infancy.  
 Moreover, as concerns sampling approaches,  
  the only methods for multilayer networks (i.e., \cite{Gjoka2011} and \cite{KhadangiBS16}) belong to the exploration-based sampling subcategory, whereas no representative exists for the random access subcategory.

Aggregation-oriented simplification category includes methods characterized by relative numerosity of methods such as multilayer community detection and, to a lesser extent, position equivalence, but also by  types of methods that are underrepresented or not represented at all. For instance, multilayer aggregation   counts only the method proposed in~\cite{DeDomenico2015} by De Domenico et al., flattening   enumerates two methods~\cite{Dickison2016,BerlingerioCG11}, whereas no multilayer extensions have been so far proposed for graph compression and coarsening.

As concerns transformation-based techniques, on the one hand, projection-based methods can be borrowed from multi-mode network context, since it is straightforward to compare different node types to nodes belonging to different layers. 
 On the other hand, the graph embedding vein experienced an unceasing and growing attention in the last few years. Nonetheless, only a handful of methods have been specifically devised for multilayer networks.  
The process of designing a graph embedding method includes several sensitive components which individually could decree the success or failure. One of the most challenging tasks lies in the way we gather structural information from the graph, e.g., by exploiting random walks, or whether we chose to use a more compact representation such as the normalized adjacency matrix. Another critical step which directly affects the amount of information retained into the learned representation is the encoding function the model aims to learn, which is in charge of leveraging the input data by encoding it in every node-embedding.  An essential component is represented by the   loss function which drives the learning process, and iteration by iteration determines the final quality of the learned representation. 
Assuming  to be able to  successfully design the above-stated components of a model, the elephant in the room is represented by scalability and the need for re-training the model in the case where even a single node is added to the graph (this issue in particular affects most of the methods for both multilayer and single-layer networks currently in the literature).

 \subsection{Future directions}  
The increasing ability of gathering and storing huge amounts of data and, consequently,  the opportunity of modeling rich-content  networks in several domains, raises the need for the development of next-generation  simplification methods, which should be   effective in leveraging  external knowledge or side-information as well as  time-based attributes when dealing with the complexity of multilayer networks.

\paragraph{Simplification for attributed networks} 
Accounting for rich-content information  that may be associated with actors, nodes, edges and/or layers represents an interesting direction to enhance the quality of a network simplification task. However, when available, such information is often sparse and incomplete,  therefore one challenge is to develop methods that can profitably exploit such information and properly embed it within the structure space, which becomes further complicated in a multilayer network setting.

\paragraph{Simplification for time-evolving networks} 
Most existing methods of multilayer network simplification deal with static graphs,   focusing on single snapshots  of the complex system while discarding the time dimension across the layers. Therefore,   effort should be made toward the development of simplification methods that are able to fully leverage the temporal dimension and the evolution of a multilayer network   over time, which might be beneficial for evolutionary as well as online/stream processing tasks.  

\paragraph{Improving interpretation of multilayer network representation feature} 
 We discussed the role of deep neural networks in representation learning problems to support transformation-oriented simplification.   
 As in other data domains, however,  deep-learning-based techniques generally fail in providing  learned representations that are highly interpretable.  
 This is further exacerbated in a multilayer network since each embedding component can be designed to capture properties related to nodes, actors, edges, and/or layers.  
 Attention should hence be devoted to advances in representation learning to develop better and more understandable embedding-based transformations of multilayer networks.

\paragraph{Emergence for evaluation benchmarks} 
Another area of improvement concerns the development of standardized evaluation benchmarks which might enable researchers and practitioners to fairly compare network simplification methods. Ultimately, this would allow us to take more robust decisions about how to choose the most appropriate method for the task at hand.

\end{document}